\documentclass{article}

\usepackage{arxiv}

\usepackage[utf8]{inputenc} 
\usepackage[T1]{fontenc}    
\usepackage{hyperref}       
\usepackage{url}            
\usepackage{booktabs}       
\usepackage{amsfonts}       
\usepackage{nicefrac}       
\usepackage{microtype}      
\usepackage{lipsum}
\usepackage{graphicx}
\usepackage{subcaption}
\usepackage{multirow}
\usepackage{nicefrac, xfrac}
\usepackage{amsmath}
\graphicspath{ {./images/} }

\title{Deploying Self-Supervised Learning for Real Seismic Data Denoising}

\author{
Giovanny A. M. Arboleda \\
  COPPE \\ 
  Federal University of Rio de Janeiro \\
  Rio de Janeiro, Brazil \\
  \texttt{arboleda@lamce.coppe.ufrj.br} \\
  \And
Claudio D. T. de Souza  \\
  COPPE \\ 
  Federal University of Rio de Janeiro \\
  Rio de Janeiro, Brazil \\
  \texttt{disouza@cos.ufrj.br} \\
  \And
Carlos E. M. dos Anjos \\
  COPPE \\ 
  Federal University of Rio de Janeiro \\
  Rio de Janeiro, Brazil \\
  \texttt{carlos.menezes@poli.ufrj.br} \\
  \And
Lessandro de S. S. Valente \\
  COPPE \\ Federal University of Rio de Janeiro \\
  Rio de Janeiro, Brazil \\
  \texttt{lessandro.sadala@gmail.com} \\
  \And
Roosevelt de L. Sardinha \\ 
  COPPE \\ 
  Federal University of Rio de Janeiro \\
  Rio de Janeiro, Brazil \\
  \texttt{roosevelt1@gmail.com} \\
  \And
Albino Aveleda \\
  COPPE \\ Federal University of Rio de Janeiro \\
  Rio de Janeiro, Brazil \\
  \texttt{bino@nacad.ufrj.br} \\ 
  \And
Pablo M. Barros \\
  CENPES, \\
  Petrobras\\
  Rio de Janeiro, Brazil \\
  \texttt{pablobarros@petrobras.com.br} \\  
  \And
André Bulcão \\
  CENPES \\ Petrobras \\ Rio de Janeiro, Brazil \\
  \texttt{bulcao@petrobras.com.br} 
  \And
Alexandre G. Evsukoff \\
  COPPE \\ Federal University of Rio de Janeiro \\
  Rio de Janeiro, Brazil \\
  \texttt{alexandre.evsukoff@coc.ufrj.br} \\
}

\begin{document}
\maketitle
\begin{abstract}
Self-supervised learning (SSL) has emerged as a promising approach to seismic data denoising as it does not require clean reference data. In this work, the deployment of the Noisy-as-Clean (NaC) method was evaluated for real seismic data denoising under controlled conditions. Two independent seismic acquisitions, each comprising noisy and filtered data, were organized into four real datasets. The NaC SSL method was adapted to add real noise to the noisy input, controlled by a parameter. An experimental protocol with ten experiments was designed to compare different strategies for deploying the NaC SSL method with the supervised learning baseline, using identical network topology and hyperparameters. The models were evaluated in terms of denoising performance, computational cost, and generalization capability. The results show that the synthetic additive white Gaussian noise (AWGN) is inadequate for the denoising of seismic data within the NaC method, and performance strongly depends on the compatibility between the injected and actual noise characteristics. Furthermore, both the characteristics of the seismic data and the noise level influence the performance of the model. Self-supervised fine-tuning on test data has improved SSL performance, whereas no such gain was observed for fine-tuning of supervised models. Finally, NaC has shown to be a simple, effective, and model-independent method that offers a feasible solution for the denoising of real seismic data.
\end{abstract}

\section{Introduction}
Offshore seismic acquisition data are inevitably contaminated by noise derived from several sources, such as sea waves, turbulence along the streamers, wind, and instruments. Noise can compromise the quality of seismic data processing and hinder the interpretation of geological features. Therefore, seismic data denoising plays a fundamental role in the seismic processing workflow.

Conventional denoising techniques have relied on explicit assumptions about signal and noise characteristics, while they consume enormous amounts of human and computational resources. Advances in deep learning (DL) for image denoising have fostered data-driven approaches for the denoising of seismic data. A comprehensive survey of the available machine learning and DL methods in seismic data processing has been presented by \cite{1}. The authors reported that denoising is one of the most common applications of DL in seismic data processing.

Early DL approaches for seismic data denoising were based on supervised learning methods, which depend on the availability of clean (filtered) data for model training. Recently, several self-supervised learning (SSL) methods have emerged for the denoising of seismic data, mostly based on their image denoising counterparts.

Self-supervised methods for image denoising have been developed by exploiting signal coherence and noise randomness to train DL models directly on noisy data. \cite{zhang_unleashing_2024} presents a recent review of the main research directions that can be identified in the literature. Blind-spot (masking-based) methods, such as Noise2Void \cite{krull_noise2void_2019}, redefine denoising as a contextual prediction problem. During training, one region of the image is masked, and the model is trained to predict it from its spatial neighbors. Transformer-based methods, such as the Denoise Transformer (DT) \cite{zhang2023self}, also use masked noisy images as input, but their computational complexity is usually high, which can limit their use. Noisy-pair methods, such as Noise2Noise (N2N) \cite{lehtinen_noise2noise_2018}, are simple and effective denoising methods. They generate pseudo-independent training pairs from the same dataset and train the network to predict one noisy realization from another. \cite{liu_trace-wise_2023} expands the methodology of spots to trace creating a blind trace network. 

Self-supervised learning methods have rapidly become a promising approach for seismic data denoising. As they do not require clean reference data, often unavailable or a time-consuming resource, they extend the applicability of DL to real-world seismic processing workflows. However, from the practitioners' point of view, the variety of approaches becomes a problem for practical use as there is a lack of time and resources to test each one and decide which is best for their application. In addition, some methods are complex, difficult to deploy, and may be used in different scenarios that are not well explored in the literature. In this context, this work presents an evaluation of SSL for real seismic data denoising in different scenarios.

Data from two seismic acquisitions, organized into two files for each, were used to define four sets of real seismic data, containing their noisy and filtered versions. Two real swell noise files were also used within the Noisy-as-Clean (NaC) \cite{xu_noisy-as-clean_2020} SSL method, which was adapted for seismic data denoising through a noise addition procedure. The NaC method was chosen as the SSL method in this work because it is simple, model-agnostic, effective, and has shown better results in preliminary tests. An experimental protocol was designed to evaluate the SSL approach in 10 experiments in five different scenarios, in which the supervised learning model was used as a baseline. All experiments use the same DL model topology, as the objective is to evaluate SSL strategies. The models were evaluated by different criteria, analyzing their performance, processing time, and ability to generalize to new data. Finally, a set of guidelines is proposed to help practitioners choose the appropriate approach based on available resources and required performance.

The paper is organized as follows. The next section presents related work on SSL methods for image denoising and their seismic data denoising counterparts. Section three presents the datasets used in this work. Section four presents the supervised and SSL approaches as well as the DL model used to implement them. Section five presents the evaluation metrics, models' results, and a critical discussion of the results to derive general guidelines for the practitioner. Section six presents the conclusions.  

\section{Related work}
\label{sec:Related_Work}

A systematic review of SSL methods for denoising seismic data is beyond the scope of this work. This section presents an overview of the main image denoising approaches and their counterparts for seismic data denoising.

Self-supervised mask-based denoising methods rely on the assumption that noise is zero-mean and statistically independent of neighboring samples, whereas the signal exhibits spatial coherence. The noise is suppressed as the model fails to reproduce the incoherent noise and converges to a coherent (clean) image. The training uses a single set of noisy images and the loss function is defined between the masked noisy sample and its predicted value. These methods differ mainly in how the mask is constructed and in how they enforce the independence between the target pixel and its prediction. In Noise2Void \cite{krull_noise2void_2019}, randomly selected pixels are replaced and predicted using a blind-spot network that explicitly excludes the center pixel from the receptive field, preventing trivial identity mapping. A similar blind-spot strategy, with architectural refinements for improved statistical modeling, is proposed by \cite{laine_high-quality_2019}. Noise2Self \cite{batson_noise2self_2019} formalizes the approach by defining structured masking schemes that guarantee the independence between the masked pixels and their input, ensuring an unbiased estimation under independent pixel noise. Subsequent works aim to relax the strict blind-spot constraint by “unblinding” the model, either by injecting auxiliary clean information or by reorganizing features across channels, thus improving the efficiency of the estimation \cite{xie_noise2same_2020, wang_blind2unblind_2022, krull_probabilistic_2020,wu_unpaired_2020}. Self2Self \cite{quan_self2self_2020} addresses data scarcity by using dropout to generate stochastic corrupted realizations of a single noisy image for training. Neighbor2Neighbor \cite{Neighbor2Neighbor_2021} leverages paired neighboring pixels under independence assumptions. Noise2Fast \cite{lequyer_fast_2022} partitions the image into complementary chessboard subsets to create mutually informative training pairs. Together, these methods differ in masking design, independence assumptions, and strategies to mitigate information loss while preserving unbiased learning.

Seismic mask-based self-supervised denoising methods follow the same independence principles but adapt masking strategies to the spatial coherence and structural redundancy of seismic data. Based on Noise2Void, \cite{birnie_potential_2021} applied a blind spot U-net to suppress random noise in seismic sections, demonstrating performance in both synthetic and field datasets. Rather than masking individual samples, \cite{fang_unsupervised_2022} constructed training pairs from neighboring seismic traces, leveraging local spatial continuity within a U-net framework to enforce statistical independence while preserving coherent events. The Neighbor2Neighbor method was adapted by proposing a sub-sampling strategy tailored to seismic grids, generating complementary subsets to attenuate random noise using a U-net architecture \cite{wang_seismic_2024}. Similarly, \cite{liu_similarity-informed_2022} explored nearby samples to synthesize auxiliary images, reinforcing structural consistency during training. \cite{pereira_pinheiro_self-supervised_2024} presents an adaptation of Noise2Self that enables effective seismic denoising from a single noisy image, achieving competitive performance in synthetic and field data. \cite{xu_s2s-wtv_2023} combined classical random-noise attenuation and weighted total variation with the Self2Self method, integrating model-driven priors and stochastic masking within a unified deep learning scheme. \cite{GAO2024105510} presents the Swin Transformer Convolutional Residual Network (SCRN) model for simultaneous denoising and interpolation of seismic data. \cite{wang_diffusion-hybrid_2025} proposed a diffusion-hybrid framework that integrates deformable convolutions and multihead self-attention mechanisms. These methods have been validated on both synthetic and field seismic datasets, showing promising denoising performance.

In noisy-pair methods, the key assumption is that the paired samples share the same latent clean signal, while their noise components are conditionally independent. Pairs are typically constructed by sub-sampling or re-corrupting, and the loss is computed directly between two noisy versions. When valid pseudo-pairs can be formed, these approaches are simpler and often provide strong denoising performance. The Noisier2Noise (Nr2N) \cite{moran_noisier2noise_2020} and Noisy-as-Clean (NaC) \cite{xu_noisy-as-clean_2020} methods use the original image as the training target and a corrupted version of it as input. The added noise must follow the same statistical distribution as the original noise, so that the network learns to suppress noise while preserving coherent signal structures. Recorrupted-to-Recorrupted (R2R) \cite{pang_recorrupted--recorrupted_2021} generates two correlated observations from a single noise realization through a deterministic transformation. \cite{mansour_zero-shot_2023} proposed a hybrid strategy that combines the Noise2Noise and Neighbor2Neighbor concepts, enabling training from a single shot acquisition.

In seismic denoising, noisy-pair strategies have been applied to both synthetic and field data. \cite{shao_noisy2noisy_2022} adopts the N2N method for pre-stack gathers, training the model on pairs of noisy data under the assumption of conditionally independent noise. The results in synthetic and real datasets show that effective attenuation is possible without clean labels when multiple noisy realizations are available. \cite{zhao_sample2sample_2023} proposed the Sample2Sample method to avoid repeated acquisitions, based on the N2N principle. It constructs adjacent paired subsamples from a single dataset using random sampling. Using local trace redundancy, the method avoids explicit noise distribution priors and independent noisy observations. \cite{ozawa_enhancing_2025} proposed a framework based on N2N, in which information was used from the common-shot and receiver domains. \cite{li_robust_2024} combined R2R pair generation with a Siamese DL architecture to enforce consistency between corrupted inputs. The approach improves robustness for both synthetic random noise and real field noise. Similarly, \cite{cheng_effective_2023} introduced a workflow for multiple types of seismic noise based on the Iterative Data Refinement (IDR) \cite{liu2022iterativedatarefinementselfsupervised} scheme and with synthetic random, demonstrating adaptability to denoise real seismic data.

\section{Dataset description}
\label{sec:Dataset_Description}
The seismic data used in this work came from two different seismic acquisitions. Table \ref{tab:table1} shows the main characteristics of the four data files. The files  $\textnormal{1A}$ and $\textnormal{1B}$ came from the first acquisition and contain $859$ and $257$ shots, respectively, and each shot was recorded by $324$ receivers with a sampling interval of $2\,\mathrm{ms}$. The files $\textnormal{2A}$ and $\textnormal{2B}$ were issued from the second acquisition and contain $1{,}666$ and $1{,}389$ shots, respectively, with $480$ receivers per shot and were also sampled at $2\,\mathrm{ms}$. All files have similar values for the signal-to-noise ratio (SNR), as shown in Table \ref{tab:table1}.

\begin{table*}[h]
\centering
\caption{Real seismic data}
\label{tab:table1}
\begin{tabular}{lllllll}
\hline\noalign{\smallskip}
File  & Shot & Number & Sampling  & Number & Trace &  SNR  \\
 & Gathers &  of Traces &  Period &  of Samples & Interval &  \\
\noalign{\smallskip}\hline\noalign{\smallskip}
1A  & $859$  & $278{,}316$ & $2$ ms & $4{,}201$ & $8.4$s & $1.822$ \\
1B  & $257$  & $83{,}268$  & $2$ ms & $4{,}201$ & $8.4$s & $2.369$\\
2A  & $1666$ & $799{,}680$ & $2$ ms & $3{,}267$ & $6.5$s & $1.716$\\
2B  & $1389$ & $666{,}720$ & $2$ ms & $3{,}267$ & $6.5$s & $7.475$\\
\noalign{\smallskip}\hline
\end{tabular}
\end{table*}

\begin{figure*}[h]
   \centering
   \begin{subfigure}[b]{0.24\textwidth}
       \includegraphics[width=\textwidth]{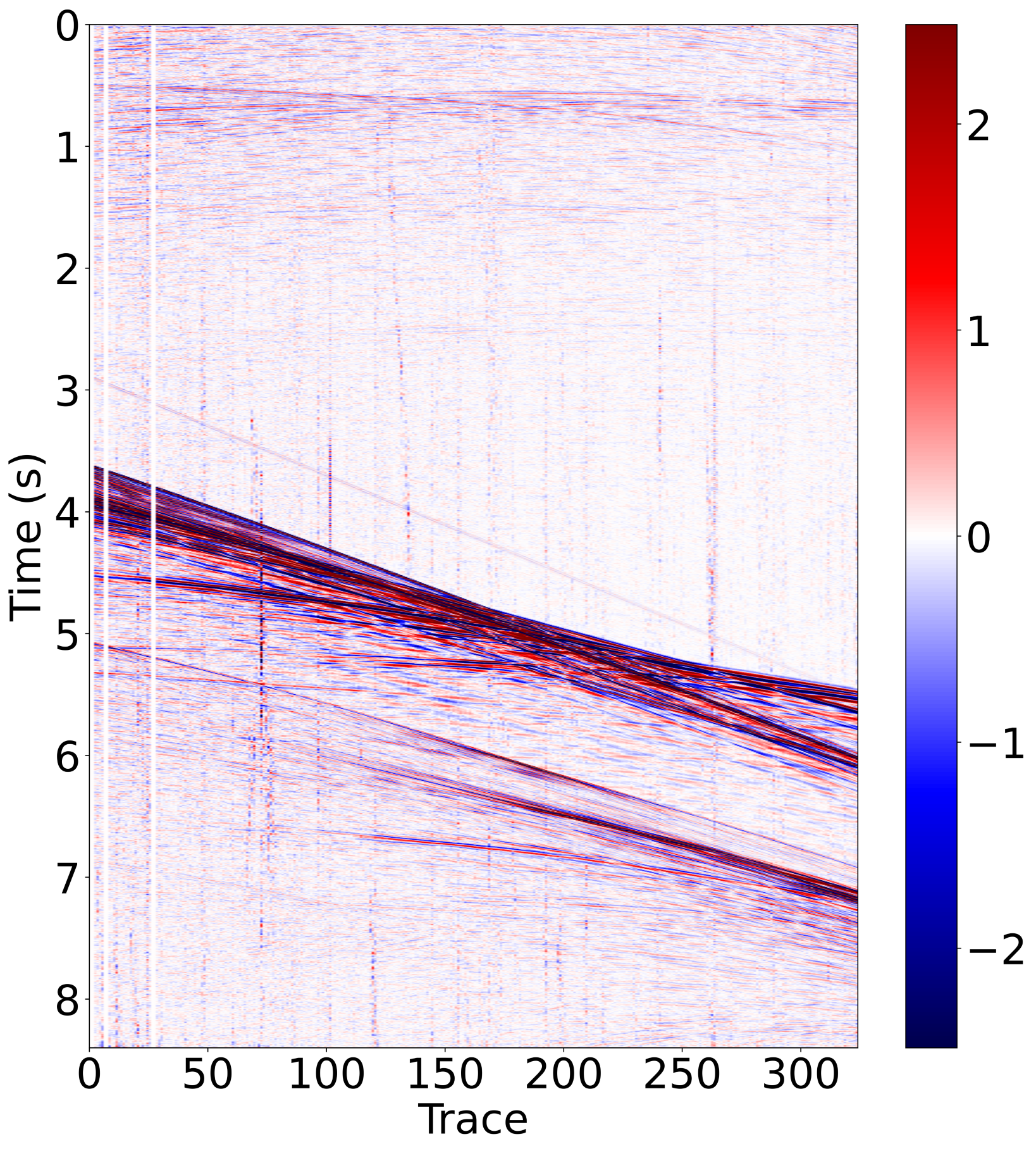}
   \end{subfigure}
   \hfill
   \begin{subfigure}[b]{0.24\textwidth}
       \includegraphics[width=\textwidth]{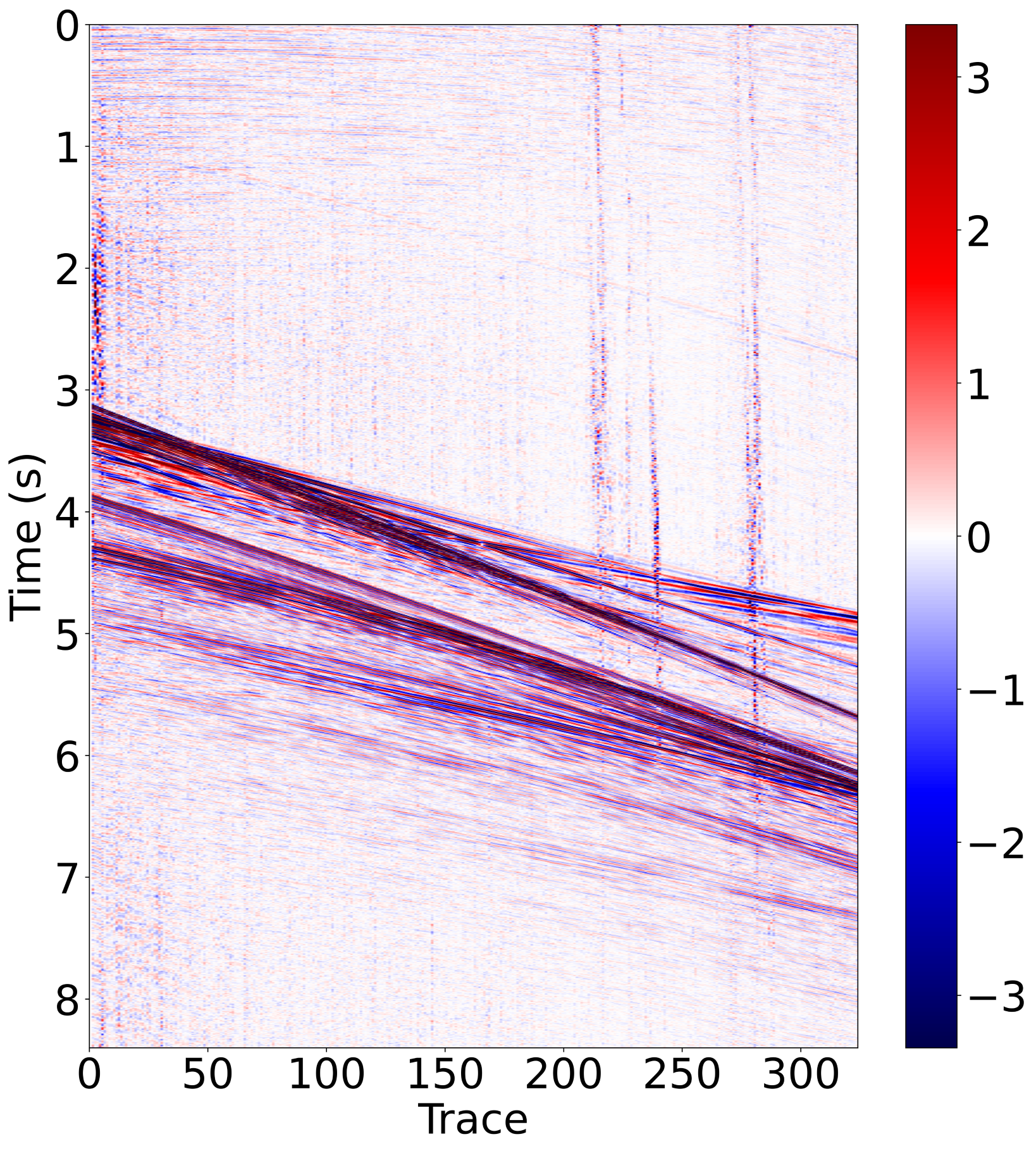}
   \end{subfigure}
   \hfill
   \begin{subfigure}[b]{0.24\textwidth}
       \includegraphics[width=\textwidth]{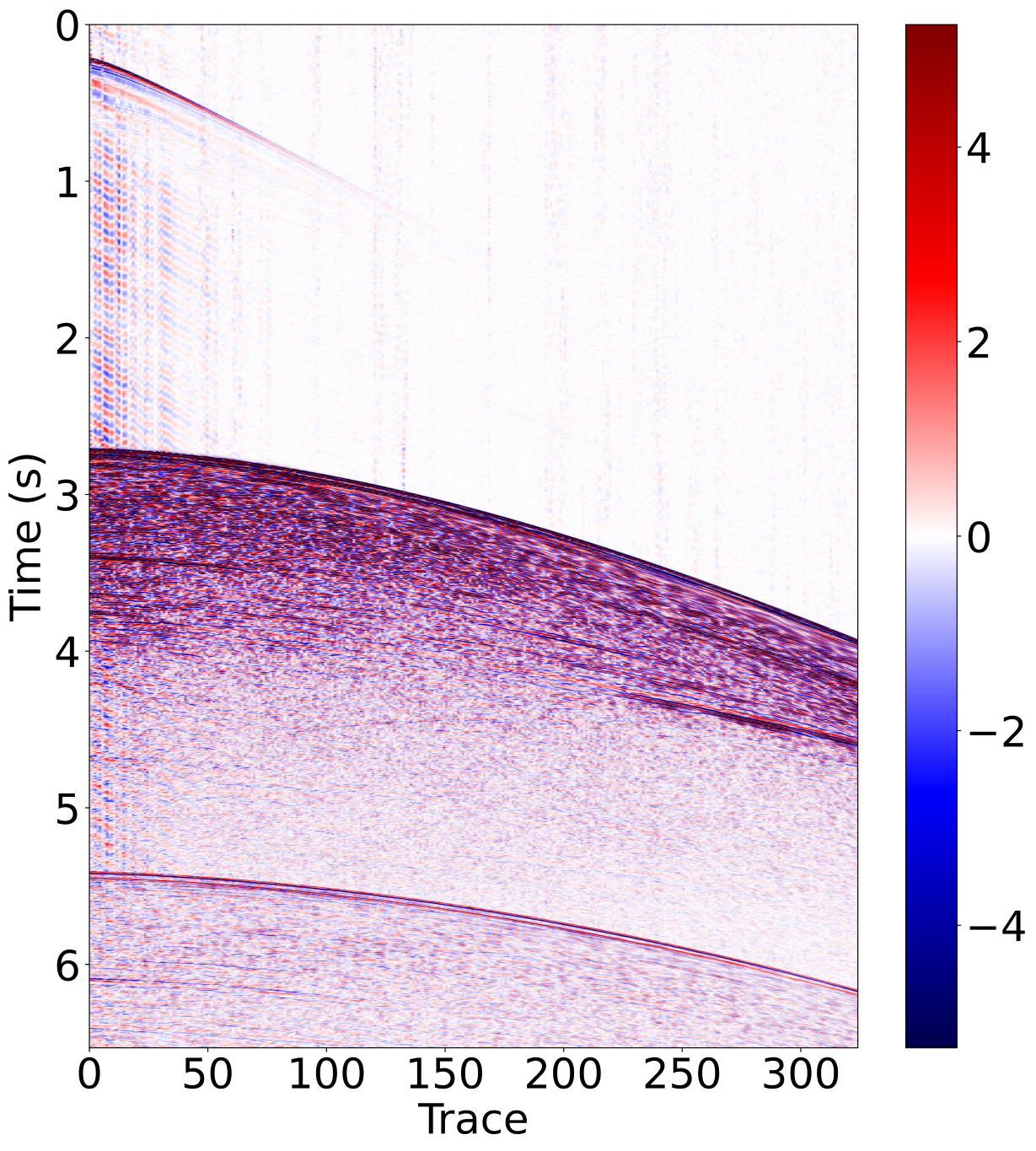}
   \end{subfigure}
   \hfill
   \begin{subfigure}[b]{0.24\textwidth}
       \includegraphics[width=\textwidth]{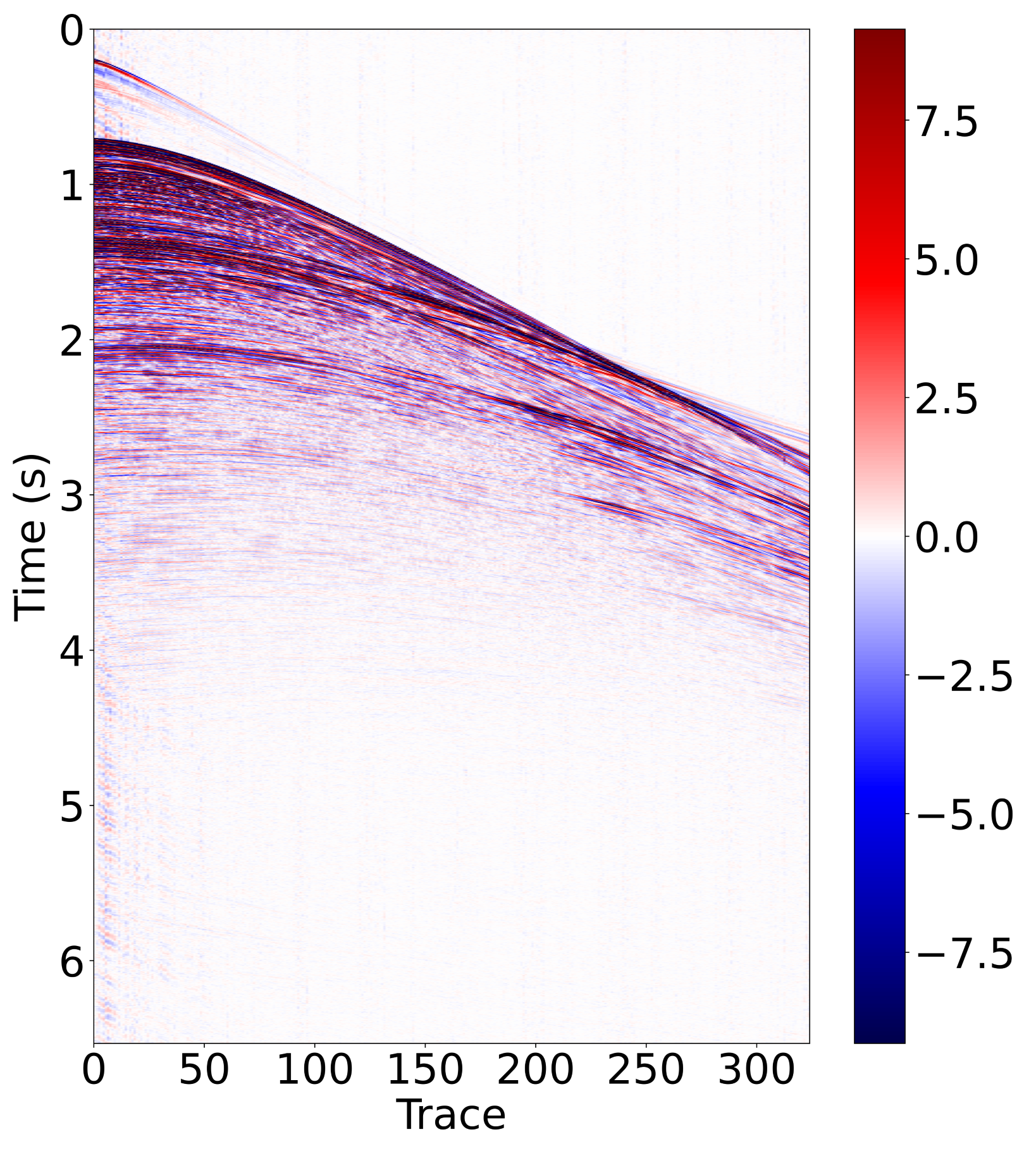}
   \end{subfigure}
   
   \vskip\baselineskip
   
   \begin{subfigure}[b]{0.24\textwidth}
       \includegraphics[width=\textwidth]{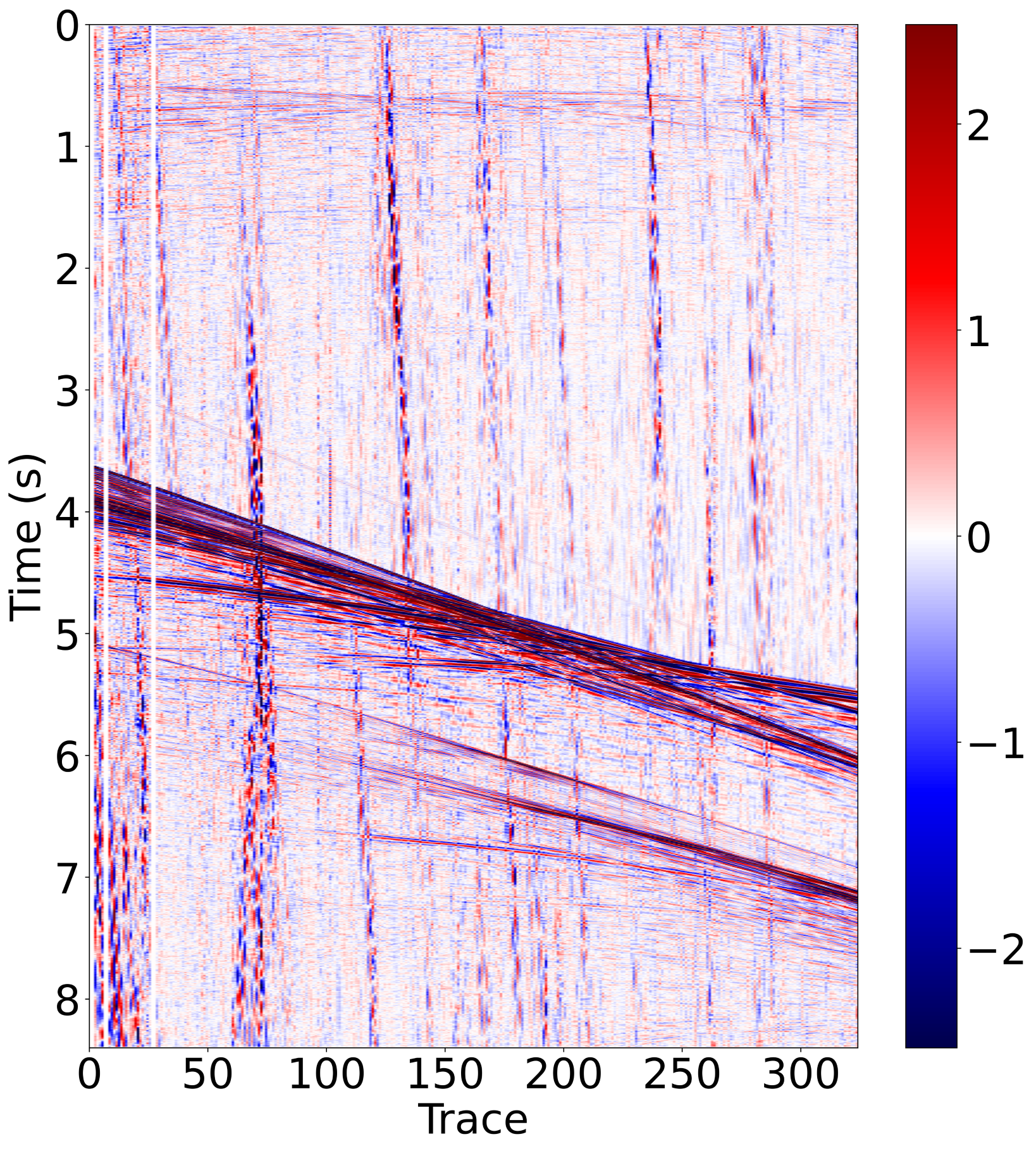}
       \caption{}
   \end{subfigure}
   \hfill
   \begin{subfigure}[b]{0.24\textwidth}
   \includegraphics[width=\textwidth]{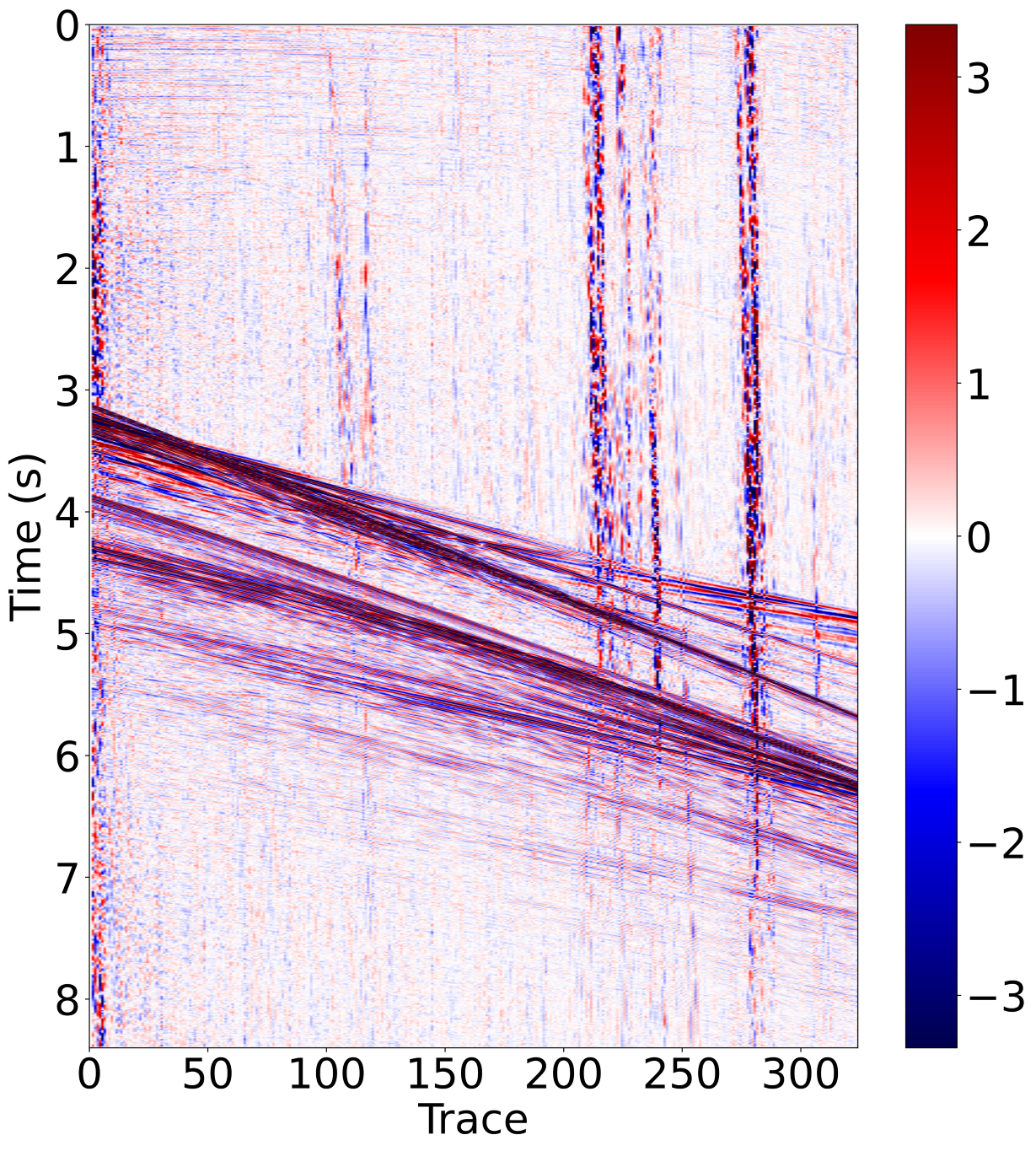}
       \caption{}
   \end{subfigure}
   \hfill
   \begin{subfigure}[b]{0.24\textwidth}
       \includegraphics[width=\textwidth]{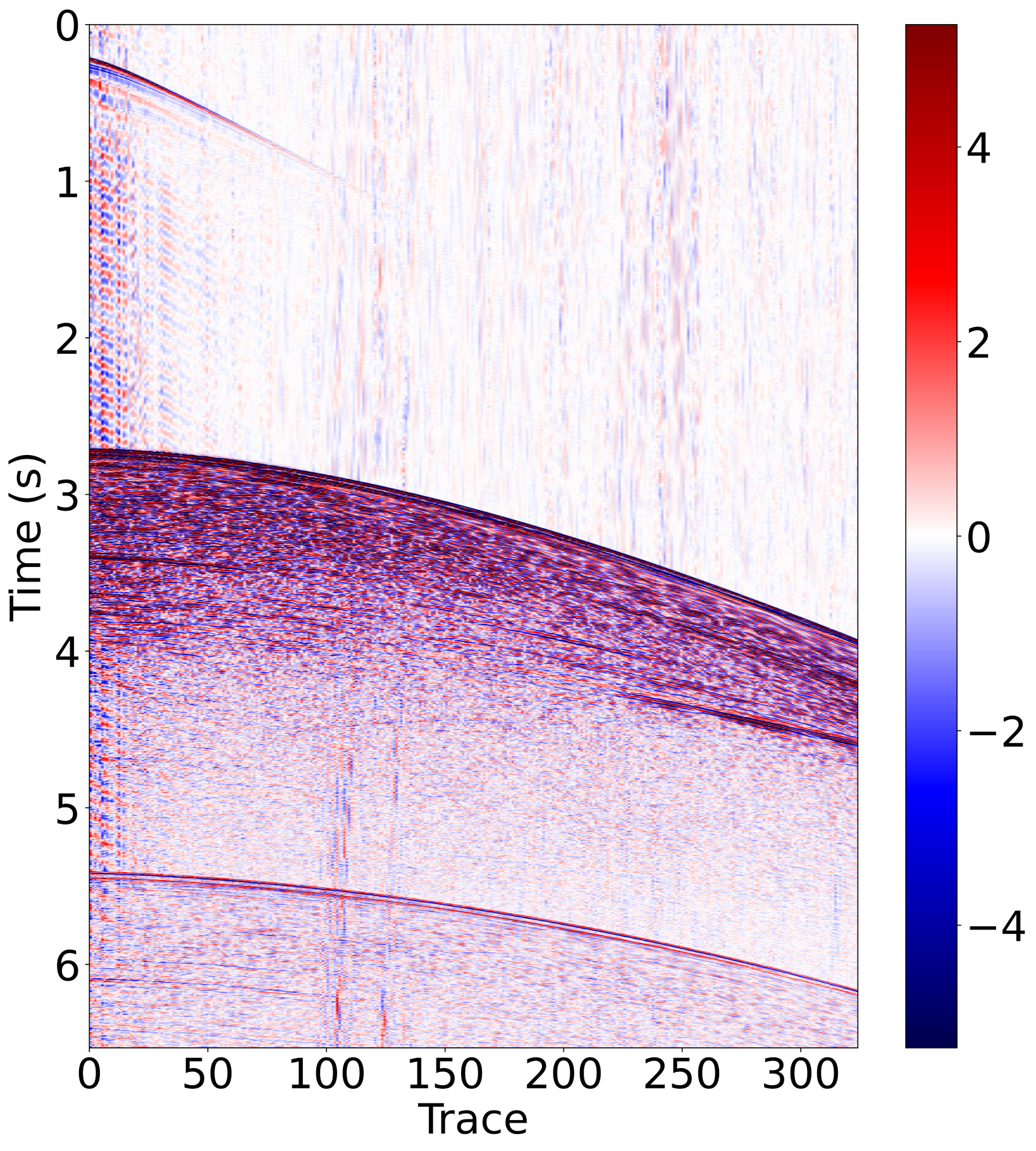}
       \caption{}
   \end{subfigure}
   \hfill
   \begin{subfigure}[b]{0.24\textwidth}
           \includegraphics[width=\textwidth]{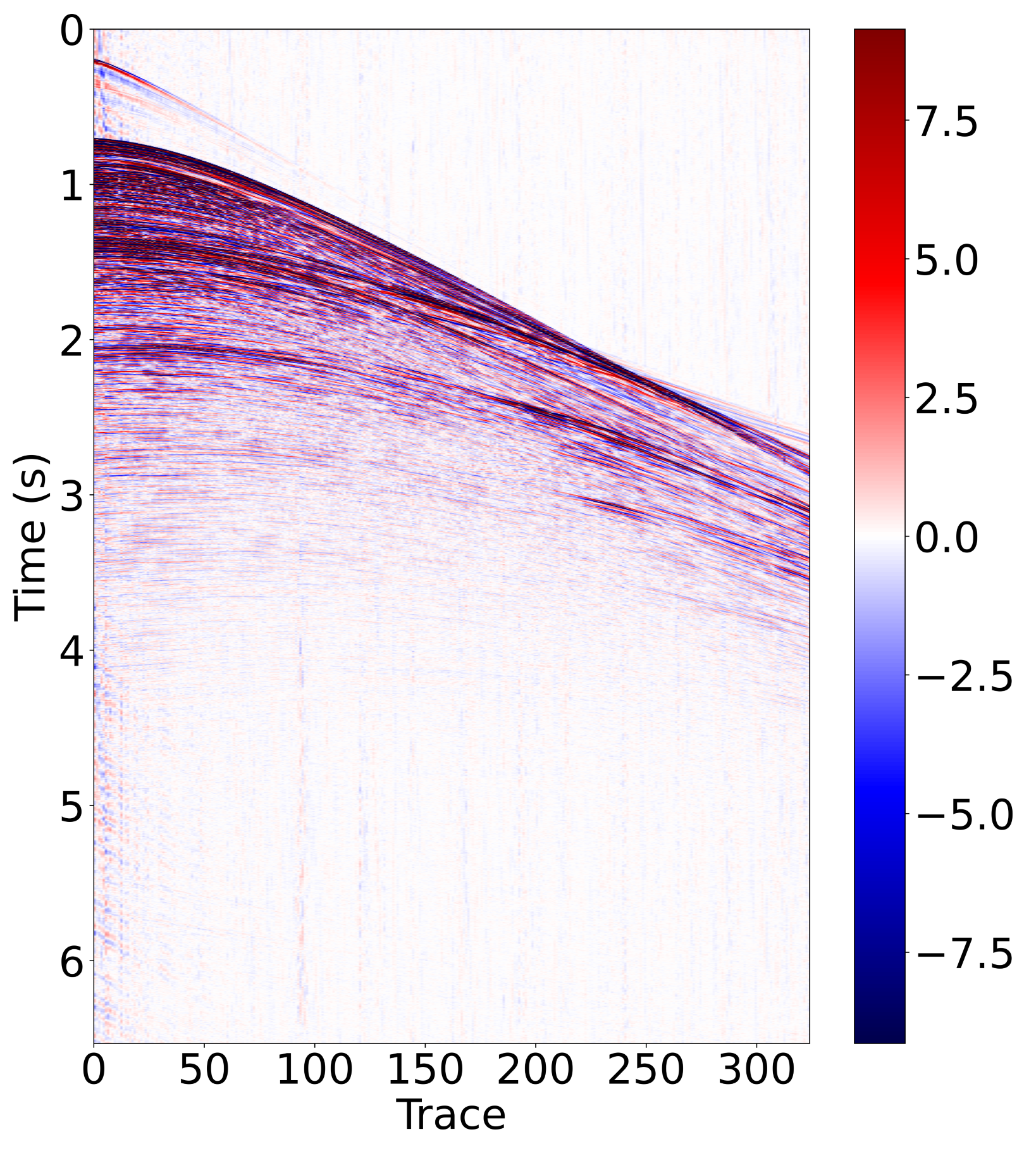}
       \caption{}
   \end{subfigure}
   \caption{Filtered (above) and Noisy (below) seismic data. (a) File 1A, (b) File 1B, (c) File 2A, and (d) File 2B.}
   \label{fig:1}
\end{figure*}

The datasets were originally provided in a noise-contaminated form, which mainly contains swell noise, as shown in Fig. \ref{fig:1}. The files were filtered by a dedicated processing workflow specifically designed for the mitigation of swell noise. Initially, a high-pass filter was implemented in the frequency domain, using transition bands of 2-3.5~Hz for files 1A and 1B and 2.5-3.5~Hz for files 2A and 2B. This step was aimed at suppressing very low-frequency components typically associated with swell noise while preserving the seismic signal of interest. Subsequently, an iterative predictive filter was applied in the $f$-$x$ domain, both in the channel and in the shot domains, taking advantage of the spatial coherence of the seismic signal in contrast to the predominantly incoherent nature of the swell noise. The algorithm was applied sequentially over multiple frequency bands, namely 0-4~Hz, 2-6~Hz, 4-8~Hz, and 6-15~Hz, with extensions up to 20~Hz for datasets 2A and 2B. This multiband strategy was adopted to capture different spectral manifestations of swell noise in the seismic bandwidth and to promote progressive attenuation while minimizing the potential degradation of the reflected signal.

The data resulting from the filtering workflow are shown in Fig. \ref{fig:1}. Although applied processing resulted in a substantial reduction in swell noise, residual noise remains present in the filtered data. In particular, datasets 1A and 1B exhibit more pronounced residual noise levels compared to datasets 2A and 2B, for which attenuation proved to be more effective. In addition, bubble noise associated with the seismic shots could not be satisfactorily removed and remains visible in the processed records.

The two additional noise files were obtained from the same two seismic acquisitions, but were extracted from files different from those used in this work, by a similar procedure as described above. The noise extraction process was designed to fully preserve the seismic signal, ensuring that the residue contains only noise. The main characteristics of the noise files are shown in  Table \ref{tab:extra1}. 

\begin{figure}[h]
   \centering
   \begin{subfigure}[b]{0.4\linewidth}
   \centering
       \includegraphics[width=\linewidth]{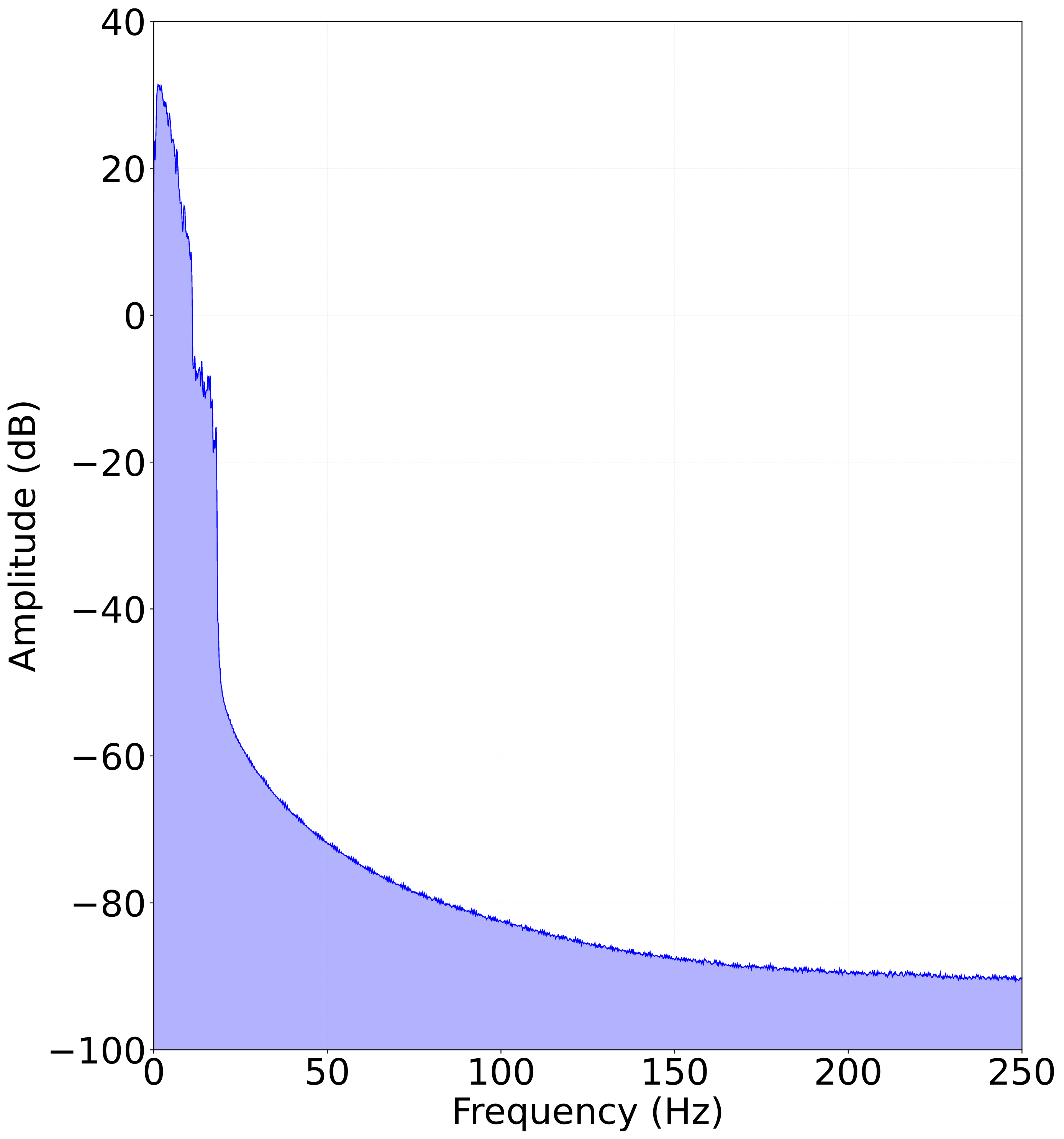}
       \caption{}
   \end{subfigure}
   \hspace{5mm}
   \begin{subfigure}[b]{0.4\linewidth}
   \centering
       \includegraphics[width=\linewidth]{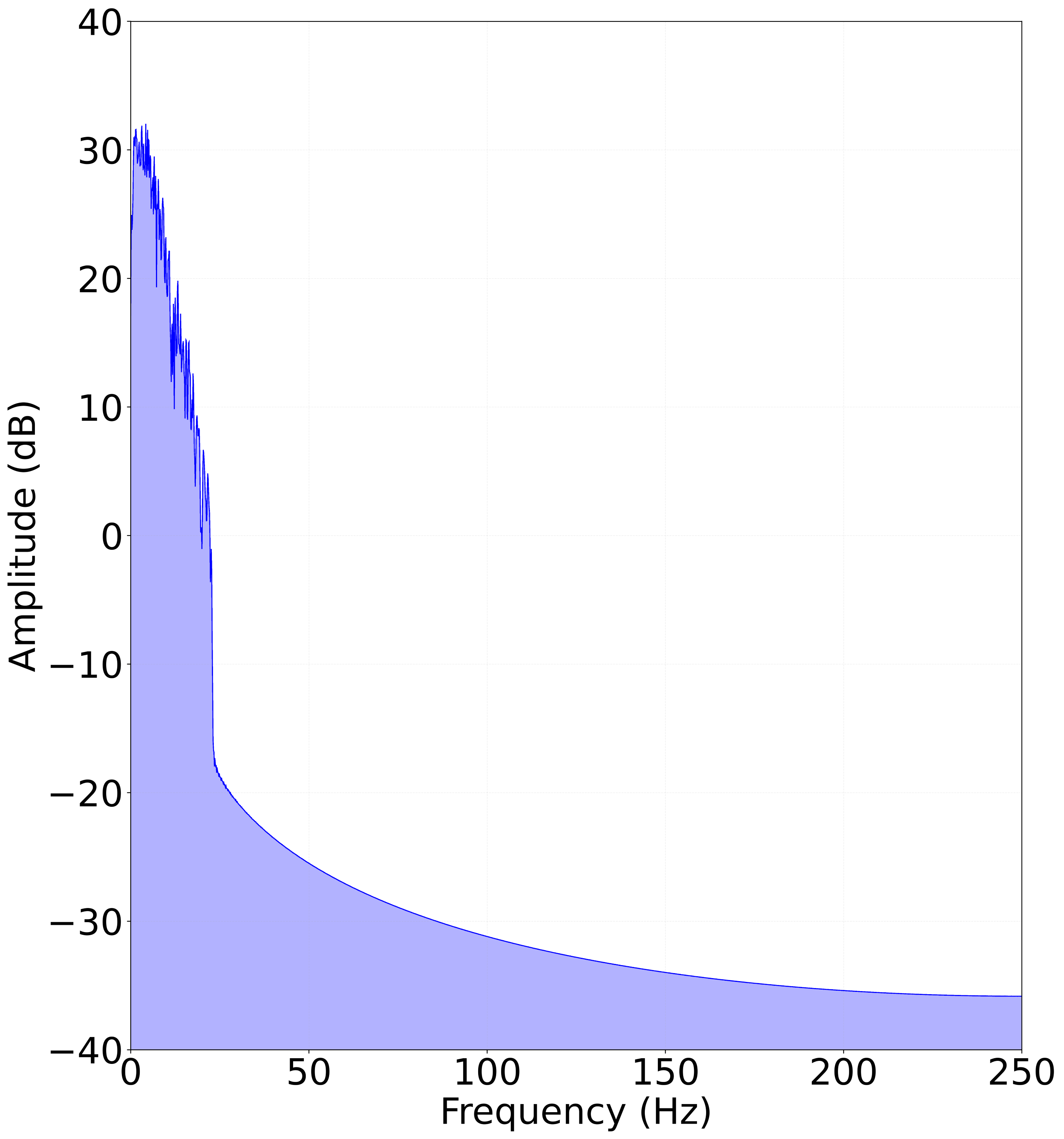}
       \caption{}
   \end{subfigure}
   \hfill
   \caption{FFT of the noise files (a) NOISE 1, (b) NOISE 2}
   \label{fig:fft_noise}
\end{figure}

The spectral characteristic of the swell noise is a well-defined peak around 5-6 Hz that decays rapidly towards higher frequencies. It has a high spatial coherence along the receiver array and low apparent velocity in the $f$–$k$ domain. The frequency spectra (FFT) of the noise files are shown in Fig. \ref{fig:fft_noise}. No significant amplitude is observed in the signal frequency range (outside the noise frequency) in the noise files, which, in fact, are very similar.

Seismic and noise data files were used to train DL models using supervised and SSL methods, as described below.

\begin{table*}[htpb]
\centering
\caption{Noise data obtained from real data by subtracting the processed data from the original raw seismic data.}
\label{tab:extra1}
\begin{tabular}{llllll}
\hline
File  & Shot & Number & Sampling  & Number & Trace  \\
 & Gathers &  of Traces &  Period &  of Samples & Interval  \\
\hline\noalign{\smallskip}
NOISE 1   & $1,904$  & $6,168,956$    & $2$ ms  & $4,201$  & $8.4$ s     \\
NOISE 2   & $1,672$  & $9,630,720$    & $2$ ms  & $3,267$  & $6.5$ s                                        \\ \hline
\end{tabular}
\end{table*}

\section{Methods}
\label{sec:Methods}
\subsection{Supervised learning} \label{subsec:Nr2N}
Swell noise in seismic data is considered to be zero mean and additive, such that it can be described as:

\begin{equation}
\label{eq:1}
   x = y + n,
\end{equation}

where $x$ represents the noisy data; $y$ represents clean data and $n$ is the observed noise.

In supervised learning, a training set composed of clean and noisy data pairs such as $\{(x_i,y_i), i=1,..,N\}$ is supposed to be available, where $x_i$ is the input to the model and $y_i$ is the target. The model is trained to compute an estimate of the clean data as  ${\hat{y}}_i=f(x_i,\theta)$, of which the optimal parameters $\theta^*$ are computed as the solution of the optimization problem:

\begin{equation}
\label{eq:2}
  \theta^* = \arg \min_{\theta}{\frac{1}{N}}\sum_{i}\mathcal{L}(f(x_i,\theta),y_i)
\end{equation}

where $\mathcal{L}$ is a loss function and $N$ the number of records in the training set.

Seismic data can vary significantly in scale, so it should be rescaled before being processed by the model. A linear scaling was applied to each shot gather of the input (noisy) data in the form of a modified version of the z-score:

\begin{equation}
\label{eq:3}
   \hat{x} = \dfrac{x - \mu}{3\sigma},
\end{equation}

where $\mu$ and $\sigma$ are, respectively, the average and standard deviation of all values in the seismic file.

After rescaling, the noisy and clean data of each shot gathered were cut into a paired set of crops with a size of $64\times64$, which were actually used to train the models. The size of the training set was defined by the total number of records $N$, where each record is a pair of input-target crops, cut at the same position. For each shot gathered, the associated crops were randomly selected until the required number of records was reached for the training set.

\subsection{Self-supervised learning} \label{subsec:NaC}
The Noisier2Noise (Nr2N) \cite{moran_noisier2noise_2020} and Noise-as-Clean (NaC) \cite{xu_noisy-as-clean_2020} methods are similar approaches that avoid the absence of clean target data by using a noisier version of the input as:

\begin{equation}
\label{eq:4}
   z = x + m,
\end{equation}

where $z=y+n+m$ represents the noisier data; $x$ represents observed noisy data, and $m$ is a zero-mean additive noise, which is assumed to be statistically close to the observed noise $n$ (cf Eq. (\ref{eq:1})).

In the NaC method for image denoising, it is assumed that the intensity of the signal is stronger than that of the noise, that is, the expectation $\mathbb{E}[x] \gg\ \mathbb{E}[n]$ and the variance $\mathbb{V}[x] \gg\ \mathbb{V}[n]$. Moreover, random variables $n$ and $m$ are considered to be issued from the same distribution such that $\mathbb{E}[n] \approx \mathbb{E}[m]$ and $\mathbb{V}[n] \approx \mathbb{V}[m]$. Under these conditions, \cite{xu_noisy-as-clean_2020} have shown that the model parameters computed as the solution of the supervised learning problem (\ref{eq:2}) will not change much with the addition of small noise $m$ and $n$ to the input and the target, respectively. Therefore, the optimal model parameters of the NaC method would be an approximation of the supervised learning ones, i.e. $\hat{\theta}^* \approx \theta^*$, and can be computed as the solution to the optimization problem:

\begin{equation}
\label{eq:5}
  \hat{\theta}^* = \arg \min_{\theta}{\frac{1}{N}}\sum_i\mathcal{L}(f(z_i,\theta),x_i\ )\
\end{equation}

where $\mathcal{L}$ is the same loss function of the problem (\ref{eq:2}) and $N$ the number of records in the training set.

Intuitively, the idea behind the NaC method is that the model learns to attenuate noise in the noisier data and since it is statistically similar to the observed noise, the model should also attenuate noise in the observed data. Note that in this work, the filtered seismic data were used as clean data for model fitting through supervised learning. In this case, supervised learning is equivalent to the NaC method, since there is residual noise in the filtered data (see Fig. \ref{fig:1} However, the noise "added" to the input of the supervised learning model is precisely the noise to be extracted.

The NaC and Nr2N methods differ in the inference phase. In the NaC approach, the model obtained by solving the optimization problem (\ref{eq:5}) is applied directly to the observed noisy data $x$ to obtain an estimate of the clean data $\hat{y}$. In the Nr2N method, the model parameters are computed by solving the same optimization problem (\ref{eq:5}), but the model is applied to the noisier data $z$ to compute an estimate of noise as $z-f(z,\theta)$, which in turn is used to estimate the clean data. The NaC method is more straightforward, while Nr2N has a more intuitive name for the SSL approach.  

An ideal seismic trace representing purely oscillatory wave propagation would have a mean close to zero, but, in practice, real seismic data are not exactly zero mean due to a number of factors. However, in this work, the seismic data for DL processing were scaled by the modified z-score (cf. Eq. (\ref{eq:3})) as well as the noise files. Therefore, $\mathbb{E}[x] \approx 0$ and $\mathbb{E}[z] \approx 0$, as well as $\mathbb{E}[n] \approx 0$ and $\mathbb{E}[m] \approx 0$.

In this work, the NaC method was considered to remain valid for the denoising of seismic data, provided that the added noise $m$ is compatible with the observed noise $n$. In addition, noise should be of low intensity compared to the signal, i.e. $\mathbb{V}[x] \gg\ \mathbb{V}[n]$ and $\mathbb{V}[z] \gg\ \mathbb{V}[m]$. Therefore, the noise added to the noisy signal must be rescaled so that the variance of the noisier data can be specified by a parameter, as described in the following.

\subsection{Noise rescaling} \label{subsec:add}
The noise data $e$ issued from the noise files NOISE 1 or NOISE 2 was normalized as:

\begin{equation}
   \label{eq:6}
   \hat{e} = \dfrac{e}{\textnormal{RMS}(e)},
\end{equation}

such that $\textnormal{RMS}(\hat{e})=1$.

Swell noise is considered a zero-mean random variable such that its variance can be estimated as the root mean square (RMS). In the adaptation of the NaC SSL method for seismic data denoising, the noise $m$ of the noisier data was computed as the normalized noise $\hat{e}$ rescaled so that $\textnormal{RMS}(m)$ can be specified as a function of $\textnormal{RMS}(x)$ by the scaling parameter $\delta > 0$:

\begin{equation}
\label{eq:7}
  m = \delta\, \textnormal{RMS}(x)\, \hat{e},
\end{equation}

such that $\textnormal{RMS}(m) = \delta\, \textnormal{RMS}(x)$.

Considering that noise $m$ is independent of the noisy signal $x$, the value of $\textnormal{RMS}(z)$ can be computed as:

\begin{equation}
\label{eq:8}
  \textnormal{RMS}(z) =\sqrt{1+\delta^2}\, \textnormal{RMS}(x) = \frac{\sqrt{1+\delta^2}}{\delta}\, \textnormal{RMS}(m),
\end{equation}

Note that $\delta > 0$ so that $\textnormal{RMS}(z) > \textnormal{RMS}(x)$.

\subsection{FCNN}
\begin{figure}[h!]
 \centering
 \includegraphics[width=\textwidth]{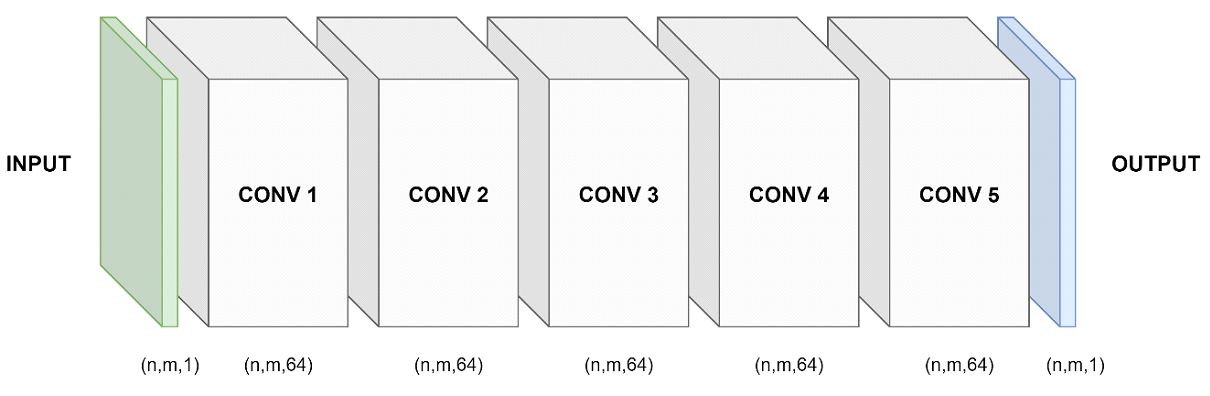}
 \caption{FCNN model with 5 layers.}
 \label{fig:5}
\end{figure}

The supervised and self-supervised learning approaches for seismic data denoising can be used with any image-to-image DL model. In this work, the Fully Convolutional Neural Network (FCNN) model was chosen because it is a lightweight model and has shown good performance in supervised seismic denoising \cite{barros_real_2024}. The FCNN is a type of CNN in which a series of convolutional layers is applied to the input \cite{17}. The main advantage of the FCNN model is that it can be applied to images of any input size. This is a useful feature for addressing problems in which the inputs have variable sizes. The 5-layer FCNN model (FCNN-5) employed in this work is presented in Fig. \ref{fig:5}, where all layers had the same size as the input. In the training phase, a crop was used as input and in the inference phase, the whole shot gather was presented to the model.

The parameters of the FCNN-5 model were adjusted by the $L_{2}$ loss function, which yielded better results in preliminary tests, and is defined as:

\begin{equation}
   \mathcal{L}(u_i,v_i) = \dfrac{1}{M}  \sum_{t \in c_{i}} (u_i(t) - v_i(t))^2,
\end{equation}

where $u_i(t)$ is the target value and $v_i(t)$ is the prediction of the model obtained for each value of the crop $c_{i}$, according to the learning approach, and $M=4.096$ is the number of values in each crop of size $64\times64$.

The topological details of FCNN-5 are shown in Table \ref{tab:table3}, where the first column represents the layer; the second column denotes the size of the kernel; the third column represents the number of kernels; and the last column represents the number of parameters in the layer. All convolutional layers use the Rectified Linear Unit (ReLU) activation function. The output layer uses 64 filters of $1\times1$ and a linear activation function. The model has a total of 1,332,678 trainable parameters. The FCNN-5 models were implemented in Python 3 with the PyTorch Lightning framework.

\begin{table}[h!]
\centering
\caption{FCNN-5 model topology}
\label{tab:table3}
\begin{tabular}{llll}
\hline\noalign{\smallskip}
Layer (Type) & Kernel Size & Kernels & Parameters \\
\noalign{\smallskip}\hline\noalign{\smallskip}
Conv2d-1  & $9\times9$  & 64    & 5,248 \\
PReLU-2   & -           & -     & 1     \\
Conv2d-3  & $9\times9$  & 64    & 331,840 \\
PReLU-4   & -           & -     & 1     \\
Conv2d-5  & $9\times9$  & 64    & 331,840 \\
PReLU-6   & -           & -     & 1     \\
Conv2d-7  & $9\times9$  & 64    & 331,840 \\
PReLU-8   & -           & -     & 1     \\
Conv2d-9  & $9\times9$  & 64    & 331,840 \\
PReLU-10  & -           & -     & 1     \\
Conv2d-11 & $1\times1$  & 64    & 65    \\
\noalign{\smallskip}\hline
\end{tabular}
\end{table}

\subsection{Experimental setup}
\label{subsec:expsetup}
Deep learning models trained with SSL are generally applied to the same data with which they were trained. However, model training typically requires high computational cost, especially for large datasets, such as in the case of seismic data denoising. Therefore, it may be useful to apply the model to new data or to fine-tune a previous model to a new problem in a low-cost SSL procedure. These scenarios have not been well explored in the literature.

The experimental setup was designed to evaluate the performance of the NaC SSL method, considering the following research questions:

\begin{itemize}
 \item \textbf{Q1}: Can the assumptions of the NaC method be validated for denoising of seismic data?
 \item \textbf{Q2}: Can the denoising performance of a model trained with SSL be similar to that of the same model trained with supervised learning?
 \item \textbf{Q3}: Can a model trained with SSL generalize to unseen data during training?
 \item \textbf{Q4}: Can SSL be used to adapt a previous model to a new problem in a low-cost fine-tuning procedure?
\end{itemize}

The above research questions were evaluated in $10$ different experiments. The experiments were divided into five scenarios, with two experiments each, using different seismic acquisitions in the training and test files (cf. \ref{tab:table_test}):

\begin{enumerate}
 \item Experiments 1 and 2: This is the pure self-supervised scheme; the models were initialized from random weights and trained and applied in the same seismic file. It was designed to evaluate SSL performance.
 \item Experiments 3 and 4: In this scenario, the self-supervised learned models were applied to a different seismic file. It was designed to evaluate SSL generalization performance.
 \item Experiments 5 and 6: In this scenario, the SSL models of the previous scenario were fine-tuned with a self-supervised procedure, using information from the test data. It was designed to evaluate whether a low-cost fine-tuning can improve the SSL generalization performance.
 \item Experiments 7 and 8: This is the pure supervised learning approach; the models were trained in one type of seismic data and applied to the other. It was designed as a baseline to compare with self-supervised approaches.
 \item Experiments 9 and 10: In this scenario, the supervised learning models were fine-tuned using information from the test data with SSL. It was designed to evaluate whether a low-cost self-supervised fine-tuning can improve the supervised learning generalization performance
\end{enumerate}

In all experiments, the models were trained under the same conditions. Noise for self-supervised training (main and/or fine-tuning) was added with a factor of $\delta = 1.0$, which achieved better results in the preliminary tests. The models were implemented using the FCNN-5 topology and trained with 1 million records (crops) in the main training phase (supervised or self-supervised) and then with 100,000 records in the fine-tuning phase. The batch size used was 64 records, the optimizer was stochastic gradient descent (SGD) with an initial learning rate of 0.01, and early stopping was implemented with a patience term of 20 epochs. The training process would be interrupted if the maximum number of epochs (200) was reached, but the training was terminated by the early stop criterion in all tests.

\begin{table*}[!htpb]
\centering
\caption{Experimental setup}
\label{tab:table_test}
\begin{tabular}{lllllll}
\hline\noalign{\smallskip}
No. & EXPERIMENT & NOISE  & INITIAL & TRAIN & NOISE & TESTING  \\
&  NAME & TRAIN & MODEL &  DATA & FINETUNE & DATA \\
\noalign{\smallskip}\hline\noalign{\smallskip}
1 & SELF(1)              & NOISE 1 & -         & \{1A, 1B\} & -       & \{1A, 1B\} \\
2 & SELF(2)              & NOISE 2 & -         & \{2A, 2B\} & -       & \{2A, 2B\} \\
\hline\noalign{\smallskip}
3 & SELF(1)-TEST(2)       & NOISE 1 & -         &  \{1A, 1B\} & -      & \{2A, 2B\} \\
4 & SELF(2)-TEST(1)       & NOISE 2 & -         &  \{2A, 2B\} & -      & \{1A, 1B\} \\
\hline\noalign{\smallskip}
5 & SELF(1)-TEST(2)-FT(2) & NOISE 1 & SELF1     & \{1A, 1B\} & NOISE 1 & \{2A, 2B\} \\
6 & SELF(2)-TEST(1)-FT(1) & NOISE 2 & SELF2     & \{2A, 2B\} & NOISE 2 & \{1A, 1B\} \\
\hline\noalign{\smallskip}
7 & SUPERVISED(1)         & -      & -         &  \{1A, 1B\} & -      & \{2A, 2B\}  \\
8 & SUPERVISED(2)         & -      & -         &  \{2A, 2B\} & -      & \{1A, 1B\}  \\
\hline\noalign{\smallskip}
9  & SUPERVISED(1)-FT(2)  & -      & SUPERVISED & \{1A, 1B\} & NOISE 1 & \{2A, 2B\} \\
10 & SUPERVISED(2)-FT(1)  & -      & SUPERVISED & \{2A, 2B\} & NOISE 2 & \{1A, 1B\} \\
\noalign{\smallskip}\hline
\end{tabular}
\end{table*}

\section{Results and Discussion}
\label{sec:Results_and_Discussion}
\subsection{Evaluation metrics}

The peak SNR (PSNR) is a commonly used metric to evaluate
image denoising methods and is defined as:

\begin{equation}
   \textnormal{PSNR} = 10 log_{10} \dfrac{\textnormal{R}^{2}}{\textnormal{MSE}},
\end{equation}

where R is a scale factor that represents the maximum value that a signal can achieve in the image representation, and
MSE is the mean squared error.

In image denoising, models are usually evaluated on 8-bit images such that R = 256. In the case of seismic data, the shot gathers are represented by real (floating point) numbers, so the maximum value in the representation does not make any sense for the PSNR calculation. In this work, a fixed scale factor of R = 2 was used in all experiments, considering that the data were rescaled according to the modified z-score method (cf. Eq. (\ref{eq:3})).

\cite{barros_real_2024} have shown that the PSNR metric alone may be limited in evaluating DL models for seismic data denoising, as it is not sensitive to signal removal and biased through low noise levels. They proposed the relative SNR ($\textnormal{SNR}^{2}$), which is computed as the ratio of the SNRs of the signal predicted by the model and the observed (noisy) signal:

\begin{equation}
   \textnormal{SNR}^{2} = 1 - \dfrac{\textnormal{SNR}(\hat{y})}{\textnormal{SNR}(x)} = 1 - \dfrac{\textnormal{RMS}(\epsilon)^2}{\textnormal{RMS}(n)^2},
\end{equation}

where $\hat{y}$ is the prediction of the model (supervised or self-supervised); $x$ is the observed data; $\epsilon = y - \hat{y}$ is the model residual and $n$ is the observed noise.

The $\textnormal{SNR}^{2}$ metric can be interpreted as the coefficient of determination $R^2$ as the relationship between the energy of the residual and the energy of the noise. The maximum value is $1.0$, but it is also possible to have negative values. This can happen when the energy of the removed noise is greater than the energy of the noise itself, indicating that some signal was removed in the denoising process.

\subsection{Results}
The results of all tests carried out in this work are shown in Table \ref{tab:table_results}.  For each experiment shown in Table \ref{tab:table_test}, two tests were performed, one for each test file. Two additional tests not described in Table \ref{tab:table_test} were included to show the performance of the NaC method using synthetic Gaussian noise. The values for PSNR and $\textnormal{SNR}^2$ metrics correspond to the average and standard deviation among all shot gathers in the test file.

The results show that additive white Gaussian noise (AWGN) is not suitable for use in the NaC method for seismic data. The pure self-supervised models (SELF) using real noise achieved comparable performance to supervised learning models (SUPERVISED), while the latter present slightly better metrics, but within the error bar defined by the standard deviation. The experiments designed to evaluate the generalization performance of self-supervised models (SELF-TEST) show that their performances are slightly worse than those of supervised models. However, the fine-tuned self-supervised models (SELF-TEST-FT) achieved the same performance as the purely self-supervised version in files 1A and 1B, and came close in file 2A. Moreover, self-supervised fine-tuning did not improve the performance of supervised learning models, which are still the same or slightly worse.

\begin{figure*}[h!]
    \centering
    \begin{subfigure}[b]{0.45\linewidth}
        \centering
        \includegraphics[width=\linewidth]{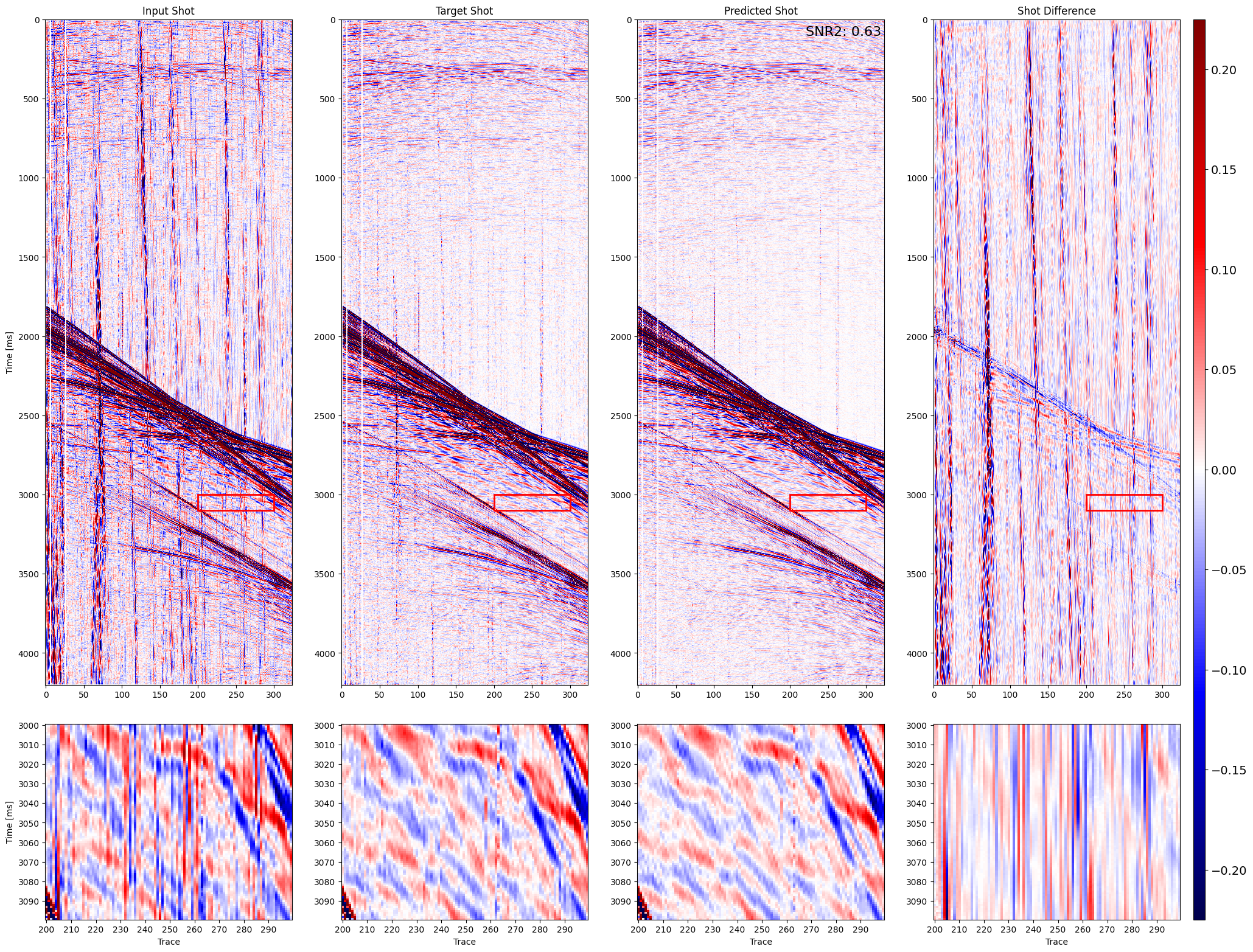}
        \caption{}
    \end{subfigure}
    \begin{subfigure}[b]{0.45\linewidth}
        \centering
        \includegraphics[width=\linewidth]{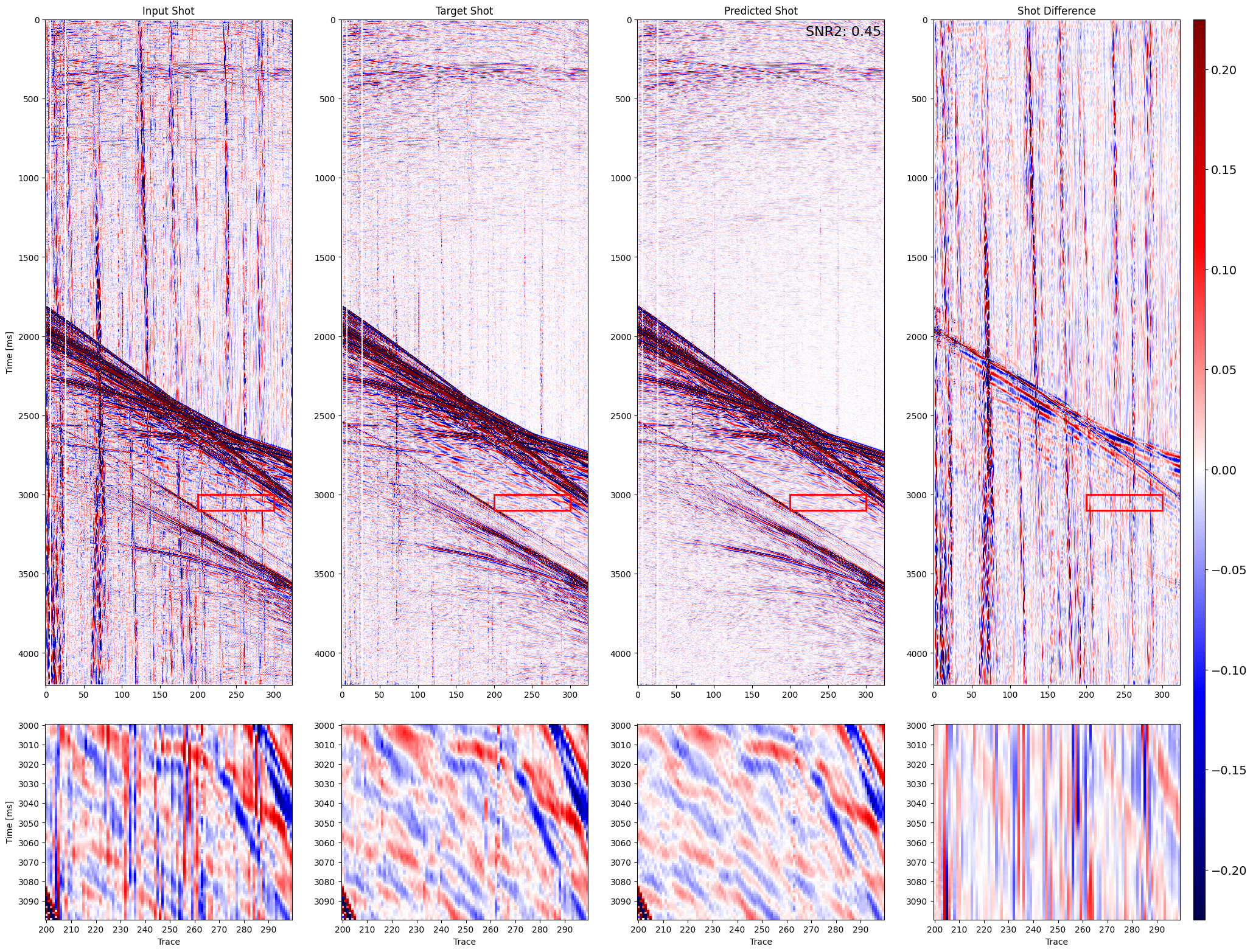}
        \caption{}
    \end{subfigure}
    \\
    \begin{subfigure}[b]{0.45\linewidth}
        \centering
        \includegraphics[width=\linewidth]{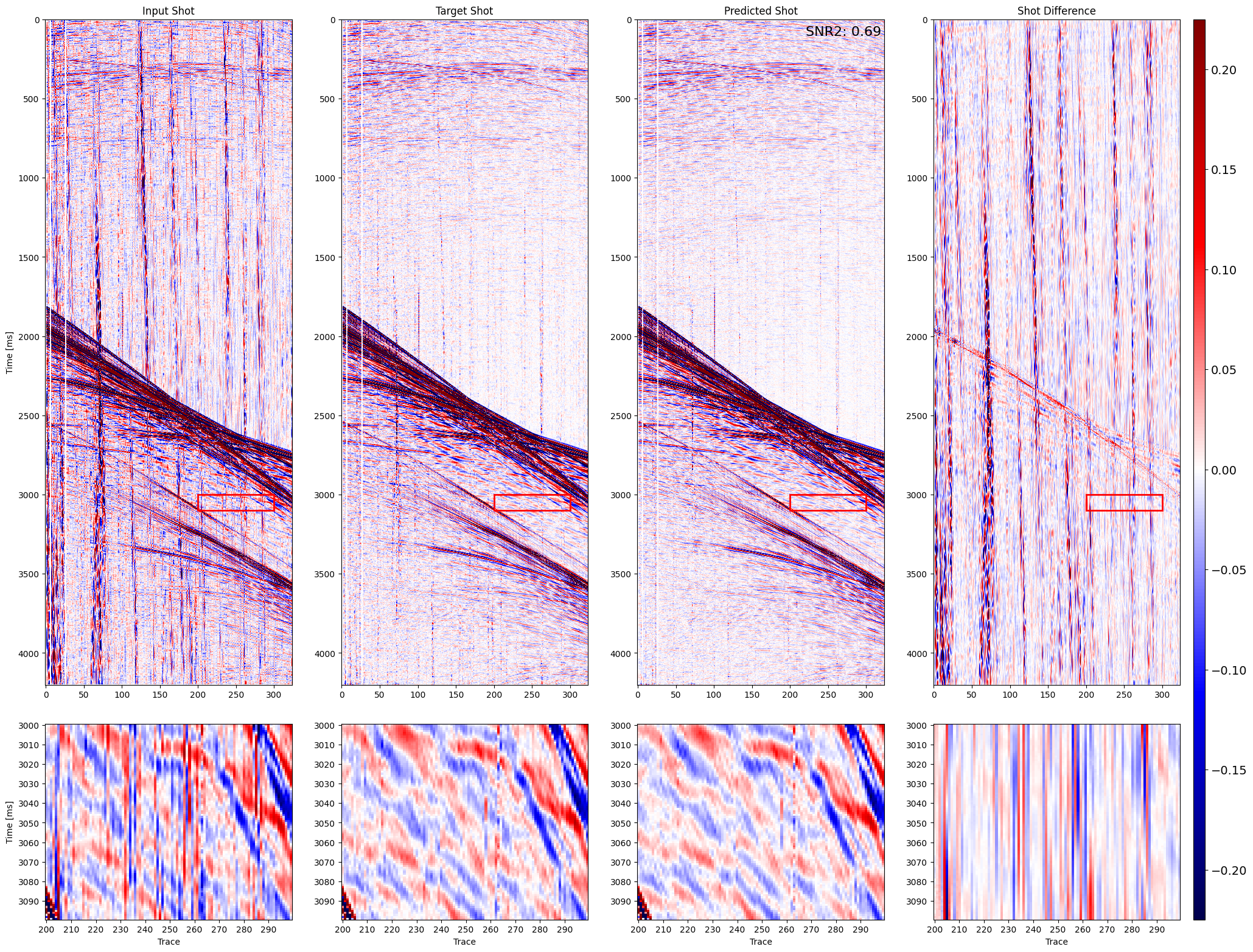}
        \caption{}
    \end{subfigure}
    \begin{subfigure}[b]{0.45\linewidth}
        \centering
        \includegraphics[width=\linewidth]{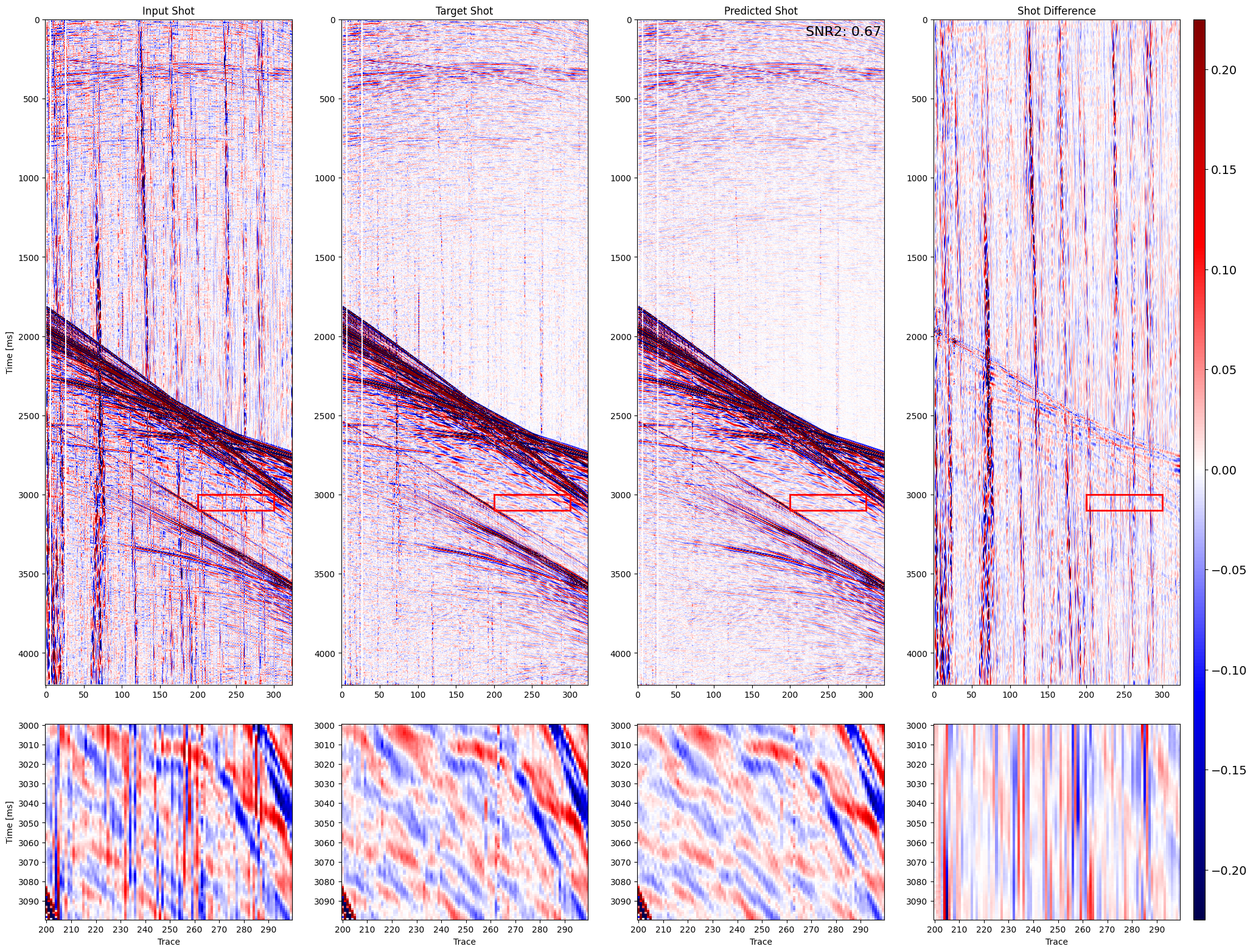}
        \caption{}
    \end{subfigure}
    \caption{Results for file data 1A: (a) SELF(1), (b) SELF(2)-TEST(1)-FT(1), (c) SUPERVISED(1), (d) SUPERVISED(2)-FT(1).}
    \label{fig:result1A}
\end{figure*}

\begin{figure*}[h!]
    \centering
    \begin{subfigure}[b]{0.45\linewidth}
        \centering
        \includegraphics[width=\linewidth]{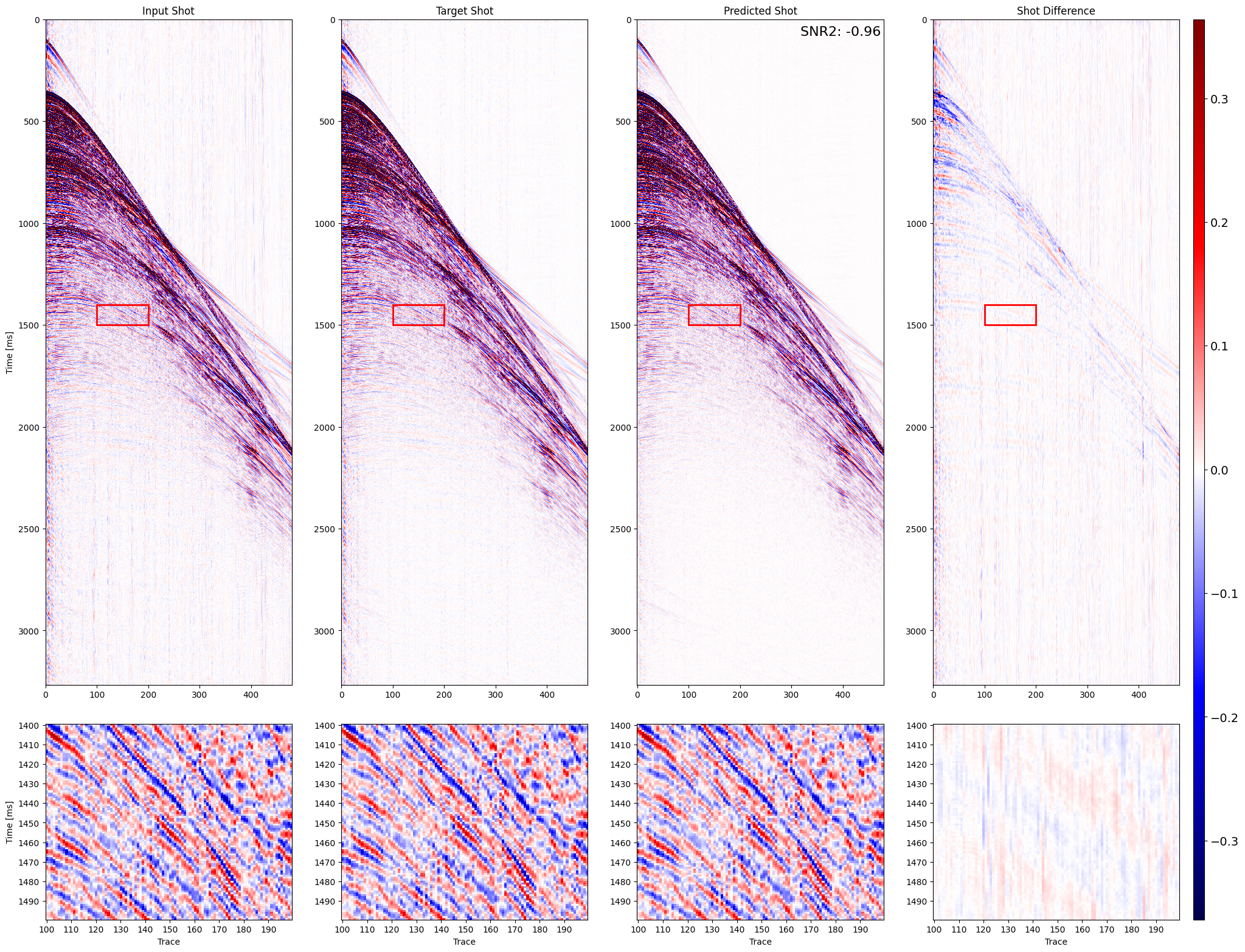}
        \caption{}
    \end{subfigure}
    \begin{subfigure}[b]{0.45\linewidth}
        \centering
        \includegraphics[width=\linewidth]{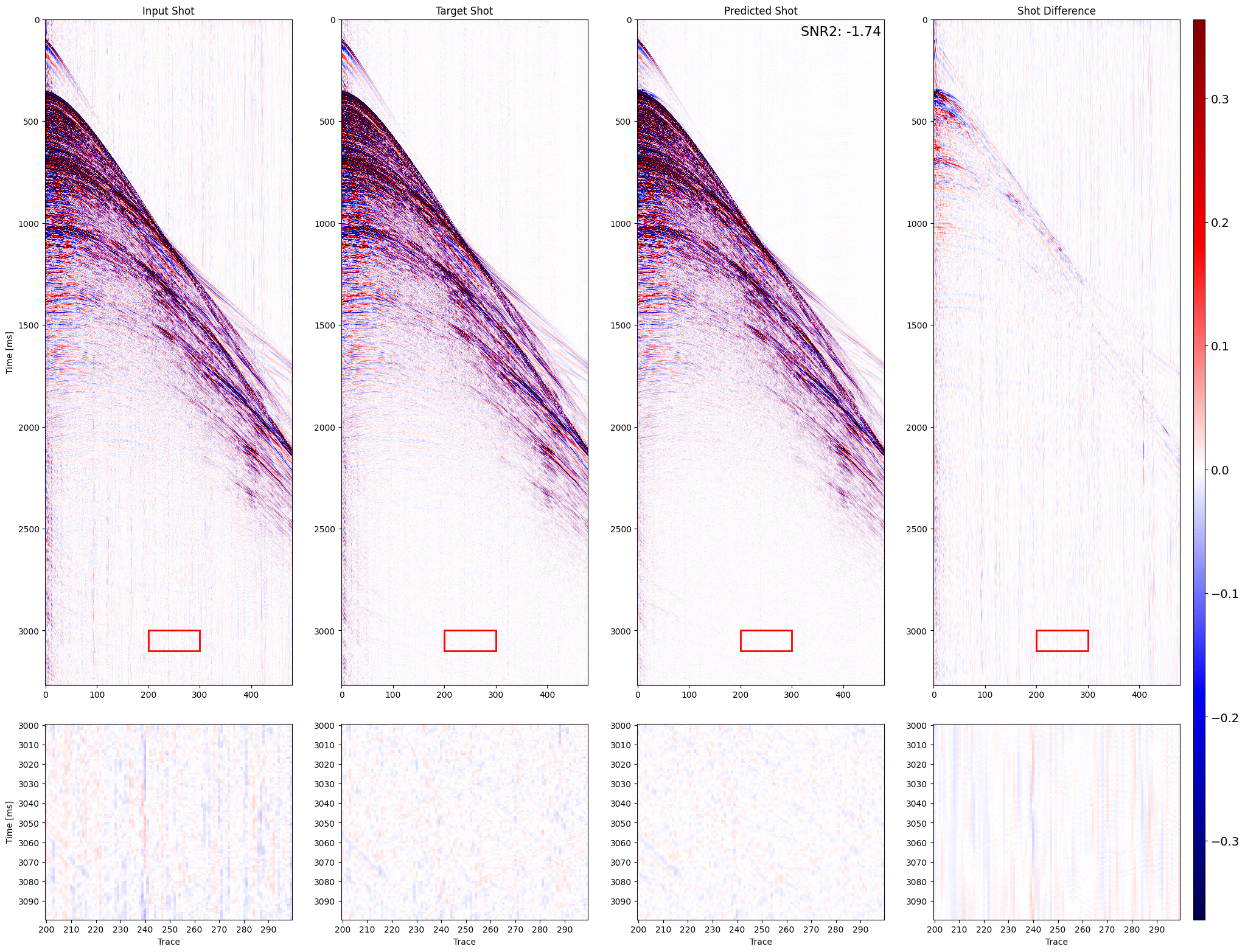}
        \caption{}
    \end{subfigure}
    \\
    \begin{subfigure}[b]{0.45\linewidth}
        \centering
        \includegraphics[width=\linewidth]{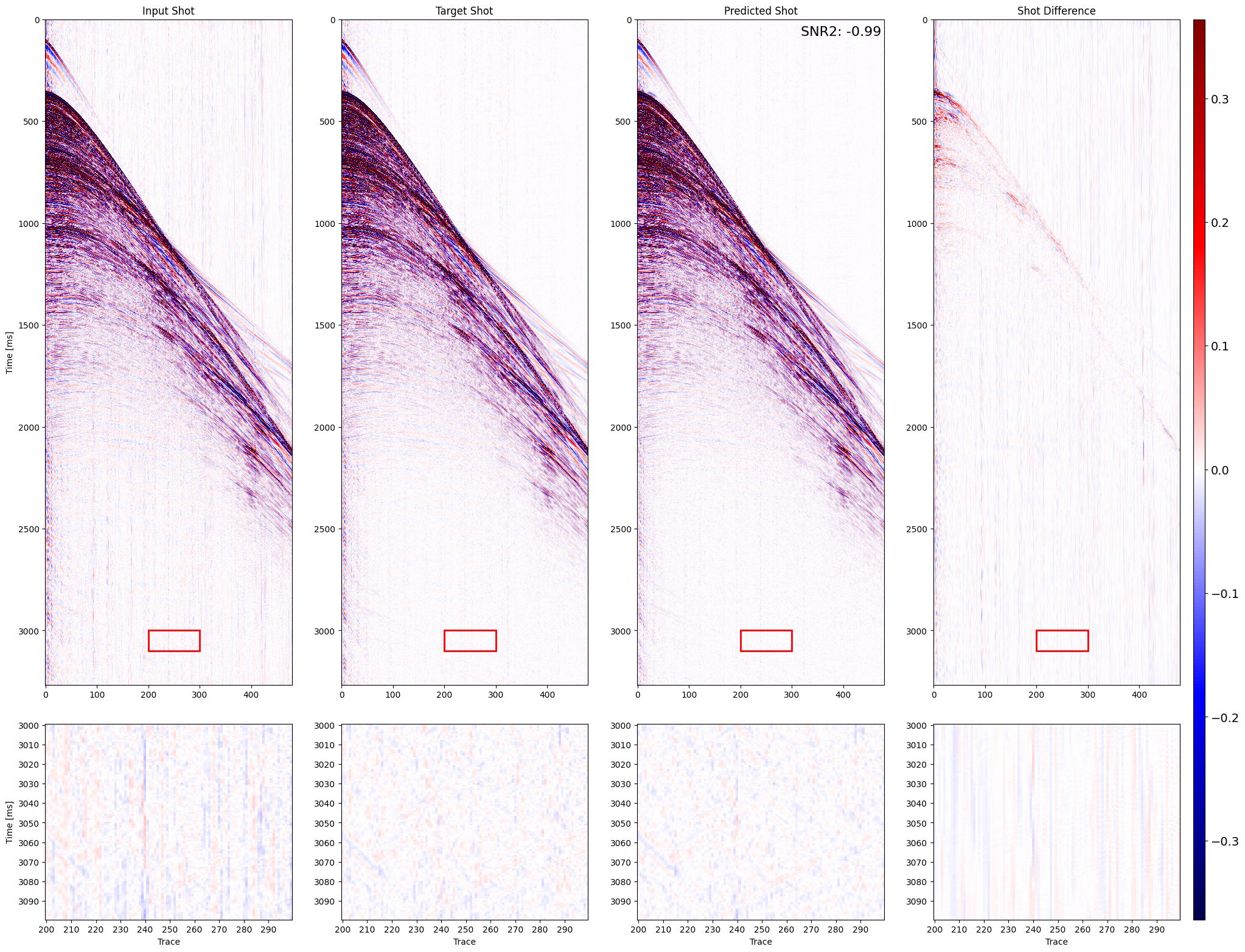}
        \caption{}
    \end{subfigure}
    \begin{subfigure}[b]{0.45\linewidth}
        \centering  \includegraphics[width=\linewidth]{Nr2N_D_BASE_LINE_fold0_shot0_Input.png}
        \caption{}
    \end{subfigure}
    \caption{Results for file data 2B: (a) SELF(2), (b) SELF(1)-TEST(2)-FT(2), (c) SUPERVISED(2), (d) SUPERVISED(1)-FT(2).}
    \label{fig:result2B}
\end{figure*}

Fig. \ref{fig:result1A} shows the results for file 1A as testing data. Each subfigure shows, from left to right, the observed noisy data, the filtered data, the model prediction, and the noise removed by the model, i.e., the difference between the noisy data and the model predictions. At the bottom of each subfigure, the region marked with a red rectangle in the main image is shown zoomed in. Fig. \ref{fig:result2B} shows the results for file 2B in the same representation. It can be seen that, although the metric values are different for different models, the results are visually very similar in both cases. In the case of file 2B (cf. Fig. \ref{fig:result2B}), the observed noisy data have a low level of noise, which has been largely removed, along with a considerable portion of the signal, which explains the negative value of $\textnormal{SNR}^2$. Moreover, a closer look at the seismogram shows that the DL models were able to remove more noise than the filter process while also removing a small portion of the signal.

All experiments were carried out with the same parameter value $\delta = 1$, as the idea was to evaluate the different strategies under the same conditions. Nevertheless, the denoising performance could be further improved by adjusting the parameter $\delta$ for each experiment. Table \ref{tab:table_results_deltas} shows the results for pure SSL under the variation of the $\delta$ parameter for the four files. The results show that the evaluation metrics were better when  $\nicefrac{1}{\delta} \approx\ \textnormal{SNR}(x)$ (cf. Table \ref{tab:table1}). Moreover, in the case of files 2A and 2B, the SSL results with adjusted $\delta$ were better than the results of the supervised baseline models.

All tests were computed on the OGBON supercomputer, in which each computing node had 360 GB of main RAM and 4 NVIDIA V100 GPUs with 32 GB of memory each. The processing times for learning and fine-tuning of the models are shown in Table \ref{tab:table_learning time}. All SSL models achieved approximately the same processing time for training. Supervised learning models also have similar training processing times for files 1A and 1B, while a longer processing time for files 2A and 2B. Recall that all models were trained using 1 million records with the same FCNN-5 topology and the same hyperparameters. The size of files 2A and 2B should not have affected the training time of the models, but the supervised models converged in a greater number of epochs in these experiments. In the fine-tuning phase, all models were trained with 100,000 records. The self-supervising models, fine-tuned with files 2A and 2B, converged is slightly longer time than the models tuned with files 1A and 1B, the same time for the fine-tuning of the supervised learning models in all examples.

Processing times for SSL can be mitigated with multi-node parallel processing. Implementing the models within the PyTorch Lightning framework allows processing on multiple nodes with minimal code modification. Experiments using 10 nodes on the OGBON supercomputer are shown in Table \ref{tab:table_10nodes}. Only pure SSL (SELF) experiments were evaluated using training data with 1 million and 10 million records. The results show that learning processing times with 1 million records drop to 5h31m for file 1A and 8h06m for file 2A. In the case of 10 million records, processing times remain the same as those of 1 million records running in one node, while no gain in denoising performance was observed. 

\begin{table*}[!htpb]
\centering
\caption{Experiments results}
\label{tab:table_results}
\begin{tabular}{llllll}
No. & EXPERIMENT & TRAINING & TESTING& PSNR &  \multicolumn{1}{l}{SNR$^2$} \\
&  NAME & DATA & DATA & & \\
\hline\noalign{\smallskip}
0   & SELF(1)-AWGN        & 1A          & 1A   & 28.399±2.715   &  \multicolumn{1}{r}{-0.155±0.196} \\
0   & SELF(2)-AWGN        & 2A          & 2A   & 28.937±1.714   & \multicolumn{1}{r}{-0.431±0.200} \\
\hline\noalign{\smallskip}
1   & SELF(1)               & 1A          & 1A   & 39.248±2.218   & \multicolumn{1}{r}{0.653±0.115} \\
4   & SELF(2)-TEST(1)       & 2A, 2B      & 1A   & 36.968±1.365   & \multicolumn{1}{r}{0.550±0.140} \\
6   & SELF(2)-TEST(1)-FT(1) & 2A, 2B      & 1A   & 38.646±1.809   & \multicolumn{1}{r}{0.630±0.114} \\
8   & SUPERVISED(2)         & 2A, 2B      & 1A   & 41.189±1.853   & \multicolumn{1}{r}{0.728±0.067} \\
10  & SUPERVISED(2)-FT(1)   & 2A, 2B      & 1A   & 40.205±1.974   & \multicolumn{1}{r}{0.693±0.083} \\
\hline\noalign{\smallskip}
1   & SELF(1)               & 1B          & 1B   & 39.268±2.219   &  \multicolumn{1}{r}{0.613±0.121} \\
4   & SELF(2)-TEST(1)       & 2A, 2B      & 1B   & 37.851±1.574   &  \multicolumn{1}{r}{0.547±0.133} \\
6   & SELF(2)-TEST(1)-FT(1) & 2A, 2B      & 1B   & 39.215±1.946   &  \multicolumn{1}{r}{0.613±0.112} \\
8   & SUPERVISED(2)         & 2A, 2B      & 1B   & 41.649±2.003   &  \multicolumn{1}{r}{0.712±0.064} \\
10  & SUPERVISED(2)-FT(1)   & 2A, 2B      & 1B   & 40.252±1.992   &  \multicolumn{1}{r}{0.659±0.091} \\
\hline\noalign{\smallskip}
2  & SELF(2)               & 2A          & 2A   & 38.800±2.102   & \multicolumn{1}{r}{0.526±0.083} \\
3  & SELF(1)-TEST(2)       & 1A,1B       & 2A   & 36.991±2.094   & \multicolumn{1}{r}{0.417±0.107} \\
5  & SELF(1)-TEST(2)-FT(2) & 1A,1B       & 2A   & 37.912±2.176   & \multicolumn{1}{r}{0.476±0.094} \\
7  & SUPERVISED(1)         & 1A,1B       & 2A   & 39.336±2.316   & \multicolumn{1}{r}{0.556±0.074} \\
9  & SUPERVISED(1)-FT(2)   & 1A,1B       & 2A   & 39.251±1.855   & \multicolumn{1}{r}{0.550±0.083} \\
\hline\noalign{\smallskip}
2  & SELF(2)               & 2B          & 2B   & 40.451±1.938   & \multicolumn{1}{r}{-0.582±0.400}  \\
3  & SELF(1)-TEST(2)       & 1A,1B       & 2B   & 37.809±2.050   & \multicolumn{1}{r}{-1.172±0.647} \\
5  & SELF(1)-TEST(2)-FT(2) & 1A,1B       & 2B   & 39.322±1.955   & \multicolumn{1}{r}{-0.809±0.490} \\
7  & SUPERVISED(1)         & 1A,1B       & 2B   & 41.010±2.285   & \multicolumn{1}{r}{-0.487±0.416} \\
9  & SUPERVISED(1)-FT(2)   & 1A,1B       & 2B   & 41.187±1.748   & \multicolumn{1}{r}{-0.458±0.381} \\
\hline\noalign{\smallskip}
\end{tabular}
\end{table*}

\begin{table*}[!htpb]
\centering
\caption{Experiments results for $\delta$ variation}
\label{tab:table_results_deltas}
\begin{tabular}{lllllll}
\hline\noalign{\smallskip}
No.& EXPERIMENT & $\delta$ & TRAINING & TESTING& PSNR &  \multicolumn{1}{l}{SNR$^2$} \\
&  NAME & & DATA & DATA & & \\
\hline\noalign{\smallskip}
1   & SELF(1) & 1              & 1A          & 1A   & 39.248±2.218   & \multicolumn{1}{r}{0.653±0.115} \\
1   & SELF(1)  & $0.5$             & 1A          & 1A   & 39.326±2.783   & \multicolumn{1}{r}{0.669±0.070} \\
1   & SELF(1)  & $0.2$             & 1A          & 1A   & 31.128±3.977   & \multicolumn{1}{r}{0.161±0.081} \\
1   & SELF(1) & $0.1$            & 1A          & 1A   & 29.885±3.381   & \multicolumn{1}{r}{0.037±0.025} \\
\hline\noalign{\smallskip}
1   & SELF(1) & 1              & 1B          & 1B   & 39.268±2.219   &  \multicolumn{1}{r}{0.613±0.121} \\
1   & SELF(1)  & $0.5$             & 1B          & 1B   & 38.560±2.404   & \multicolumn{1}{r}{0.613±0.097} \\
1   & SELF(1)  & $0.2$             & 1B          & 1B   & 31.944±4.155   & \multicolumn{1}{r}{0.190±0.067} \\
1   & SELF(1)  & $0.1$            & 1B          & 1B   & 30.488±3.717   & \multicolumn{1}{r}{0.045±0.021} \\
\hline\noalign{\smallskip}
2  & SELF(2) & 1              & 2A          & 2A   & 38.800±2.102   & \multicolumn{1}{r}{0.526±0.083} \\
2   & SELF(2)  & $0.5$             & 2A          & 2A   & 39.556±2.439   & \multicolumn{1}{r}{0.578±0.060} \\
2   & SELF(2)  & $0.2$             & 2A          & 2A   & 38.747±2.220   & \multicolumn{1}{r}{0.538±0.057} \\
2   & SELF(2)  & $0.1$            & 2A          & 2A   & 33.574±3.048   & \multicolumn{1}{r}{0.167±0.047} \\
\hline\noalign{\smallskip}
2  & SELF(2) & 1              & 2B          & 2B   & 40.451±1.938   & \multicolumn{1}{r}{-0.582±0.400}  \\
2   & SELF(2)  & $0.5$             & 2B          & 2B   & 41.958±2.523   & \multicolumn{1}{r}{-0.272±0.300} \\
2   & SELF(2)  & $0.2$             & 2B          & 2B   & 44.105±2.714   & \multicolumn{1}{r}{-0.009±0.022} \\
2   & SELF(2)  & $0.1$            & 2B          & 2B   & 44.242±2.832   & \multicolumn{1}{r}{0.028±0.214} \\
\hline\noalign{\smallskip}
\end{tabular}
\end{table*}

\begin{table*}[!htpb]
\centering
\caption{Processing time for model learning and fine-tuning with 1 node}
\label{tab:table_learning time}
\begin{tabular}{llllll}
\hline\noalign{\smallskip}
No. & EXPERIMENT & TRAINING & FINE-TUN. & PROC.TIME & PROC.TIME\\
 & NAME       & DATA & DATA & TRAINING & FINE-TUN \\
\hline\noalign{\smallskip}
1             & SELF(1)               & 1A          & -   & 1d09h43m  &  -     \\
2             & SELF(2)               & 2A          & -   & 1d13h04m  &  -     \\
\hline\noalign{\smallskip}
3             & SELF(1)-TEST(2)       & 1A,1B       & -   &  1d19h33m &  -     \\
4             & SELF(2)-TEST(1)       & 2A, 2B      & -   &  1d17h37m &  -     \\
\hline\noalign{\smallskip}
5             & SELF(1)-TEST(2)-FT(2) & 1A,1B       & 2A   & 1d19h33m &  2h06m  \\
5             & SELF(1)-TEST(2)-FT(2) & 1A,1B       & 2B   & 1d19h33m &  2h06m  \\
6             & SELF(2)-TEST(1)-FT(1) & 2A, 2B      & 1A   & 1d17h37m &  1h50m  \\
6             & SELF(2)-TEST(1)-FT(1) & 2A, 2B      & 1B   & 1d17h37m &  1h50m  \\
\hline\noalign{\smallskip}
7             & SUPERVISED(1)         & 1A,1B       & -  & 1d17h31m &  -      \\
8             & SUPERVISED(2)         & 2A, 2B      & -   & 2d08h37m &  -      \\
\hline\noalign{\smallskip}
9             & SUPERVISED(1)-FT(2)   & 1A,1B       & 2A   & 1d17h37m &  2h33m   \\
9             & SUPERVISED(1)-FT(2)   & 1A,1B       & 2B   & 1d17h37m &  2h33m   \\
10            & SUPERVISED(2)-FT(1)   & 2A, 2B      & 1A   & 2d08h37m &  2h16m   \\
10            & SUPERVISED(2)-FT(1)   & 2A, 2B      & 1B   & 2d08h37m &  2h16m   \\
\hline\noalign{\smallskip}
\end{tabular}
\end{table*}

\begin{table*}[!htpb]
\centering
\caption{Processing time for model learning with 10 nodes}
\label{tab:table_10nodes}
\begin{tabular}{lllllll}
\hline\noalign{\smallskip}
No. & EXP. & TRAINING  & RECORDS & PROC.& PSNR         & SNR$^2$        \\
 & NAME & DATA & & TIME &  &       \\
\hline\noalign{\smallskip}
1  & SELF(1)     & 1A            & 1M      & 5h31m      & 39.248±2.218 & 0.653±0.115 \\
2  & SELF(2)     & 2A            & 1M      & 8h06m      & 38.800±2.102 & 0.526±0.083 \\
\hline\noalign{\smallskip}
1  & SELF(1)     & 1A            & 10M     & 1d11h30m   & 39.273±2.215 & 0.654±0.113 \\
2  & SELF(2)     & 2A            & 10M     & 1d15h48m   & 38.653±2.064 & 0.518±0.089 \\
\hline\noalign{\smallskip}
\end{tabular}
\end{table*}

\subsection{Discussion}
The results presented in the previous section show that noise quality is of paramount importance to the performance of the NaC method. Synthetic AWGN noise is not suitable for denoising of seismic data, and the better the quality of the added real noise data, the better the results will be.

In this work, the supervised learning model can be seen as the NaC method using the best noise sample possible: the same noise and the same intensity that should be removed. In this case, the model shows good generalization performance in all tests. The type of seismic data and the noise level also affect the performance of the DL models. In the case of SSL models, self-supervised fine-tuning using information on the test data improved the performance of the model, while it was not the case for the supervised learning models.

The processing times for the models were similar for all SSL models and for the supervised learning model that used files 1A and 1B in the main training phase. The supervised learning models trained with files 2A and 2B showed twice the training time. However, in the supervised learning approach, the same model could, in principle, be used for different test files, and the inference times are much shorter than the training times. In the case of pure SSL, training would have to be performed for each test file, so training times for very large seismic files could become a problem, which could be mitigated with multi-node parallel processing. Even so, self-supervised fine-tuning works well and can be performed in a reasonable time, being an alternative solution if a previous model is available.
We can thus answer the research questions proposed in the subsection \ref{subsec:expsetup}.

\begin{itemize}
 \item [] \textbf{Q1}: Can the assumptions of the NaC method be validated for denoising of seismic data?
\end{itemize}
The main assumption of the NaC method about the compatibility of the noise to be added to the noisier data was validated, as the results show that the better the quality of the added real noise data, the better the results will be.

In the case of expectation and variance assumptions, further studies are necessary. The expectation assumption does not match for the case of seismic data, as they were pre-processed for zero mean. The assumption on the variance of the observed data  $\mathbb{V}[x] \gg\ \mathbb{V}[n]$ was partially matched only for file 2B, as shown by the SNR values in Table \ref{tab:table1}. Moreover, Table \ref{tab:table_results_deltas} shows that better results were obtained when $\nicefrac{1}{\delta} \approx \textnormal{SNR}(x)$, which implies that $\mathbb{V}[x] \approx \sqrt{\nicefrac{1+ \delta}{\delta}}\: \mathbb{V}[n]$. In the case of noisier data, the variance assumption $\mathbb{V}[z] \gg\ \mathbb{V}[m]$ was also partially matched as, by definition, $\mathbb{V}[z] = \nicefrac{\sqrt{1+\delta^2}}{\delta}\, \mathbb{V}[m]$ (cf. Eq. \ref{eq:8}).

\begin{itemize}
 \item [] \textbf{Q2}: Can the denoising performance of a model trained with SSL be similar to that of the same model trained with supervised learning?
\end{itemize}

Although the assumptions of the NaC method were not fully met, the results show that the performances of the SSL models were similar to those of the supervised learning models in most test files. Moreover, the denoising performance of SSL models with adjusted $\delta$ was improved for all files and slightly better than the supervised baseline models in the case of files 2A and 2B. Nevertheless, the processing time for self-supervised training is much greater than the processing time for inference of the supervised learning model.

\begin{itemize}  
 \item [] \textbf{Q3}: Can a model trained with SSL generalize to unseen data during training?
\end{itemize}

The generalization performance of the pure self-supervised model is not as good as the supervised model, but it can be improved with a low-cost self-supervised fine-tuning phase.

\begin{itemize}  
 \item []\textbf{Q4}: Can SSL be used to adapt a previous model to a new problem in a low-cost fine-tuning procedure?
\end{itemize}

The SSL fine-tuning approach works well in the case of a previous model also trained with SSL, but it did not improve the performance of a model trained with supervised learning.  

It is important to note that the NaC method is not fully blind as it needs a sample of the noise that is compatible with the one present in the data, and the better the compatibility of the added noise, the better the performance of the model. However, one can consider that extracting a noise sample could be more feasible than processing all data to obtain a version of the clean data for the supervised learning approach. Moreover, the NaC SSL approach could be directly applied when passive data is available, as in the case of permanent reservoir monitoring (PRM). If a curated noisy-clean dataset is available, supervised learning should be a better choice than the NaC approach for seismic data denoising. Otherwise, if a supervised training dataset is not available, the NaC method can be seen as an alternative that achieves results comparable to supervised learning with reasonable computational cost.

\section{Conclusion}
\label{sec:Conclusion}

This work has presented the deployment of the Noisy-as-Clean (NaC) method for real seismic data denoising, where real swell noise files, rather than synthetic noise, were used to generate the noisier input. Two independent seismic acquisition data, each organized into two files that contained noisy and filtered data, were used to define four real datasets for the experiments. An experimental protocol comprising ten experiments was designed to evaluate the SSL approach against the supervised learning baseline. All experiments used the same DL model topology and identical hyperparameters to ensure a controlled comparison. The models were evaluated according to different criteria,  analyzing their denoising performance, processing time, and their ability to generalize to new data.

The results have shown that the use of synthetic additive white Gaussian noise (AWGN) is not suitable for denoising seismic data within the NaC method, and the performance greatly depends on the compatibility of the real noise used to generate the noisier input. The characteristics of the seismic data and the noise level also affect the performance of DL models. In the case of SSL models, self-supervised fine-tuning using information on the test data improved the performance of the model, while it was not the case for the supervised learning model. Furthermore, the energy of the noisier input, controlled by the parameter $\delta$, can also affect the performance of the SSL model. Adjusting this parameter can improve the performance of the SSL model and even outperform the supervised model in some cases.

Although NaC does not require clean targets, it cannot be considered a fully blind denoising method. Its effectiveness depends on the availability of representative noise samples. The closer this compatibility, the better the denoising performance. Nevertheless, in practical seismic workflows, extracting representative noise samples may be more feasible than generating sufficiently clean data for supervised training. In addition to being effective, the NaC method is simple and model-agnostic. The noisier input can be processed offline, and the method can be deployed by any image-to-image DL model. In this sense, NaC offers a pragmatic compromise between strict supervision and fully blind denoising, providing a viable alternative for real-data seismic applications where clean references are unavailable.

\section*{Acknowledgements}
\label{sec:Acknowledgements}
The authors thank Petrobras for their financial support, for providing the real swell noise data used in this work, and for providing the high-performance computing environment used to run the experiments. The authors also thank the Brazilian Research Council, CNPq, and the Rio de Janeiro State Research Agency, FAPERJ, for their financial support for this research.

\section*{Declaration of competing interest}
\label{sec:Interest}
The authors declare that they have no known competing financial interests or personal relationships that could have appeared to influence the work reported in this paper

\bibliographystyle{unsrt}  
\bibliography{bibliography}  






\end{document}